\documentclass[fleqn,usenatbib]{mnras}

\usepackage[T1]{fontenc}

\DeclareRobustCommand{\VAN}[3]{#2}
\let\VANthebibliography\thebibliography
\def\thebibliography{\DeclareRobustCommand{\VAN}[3]{##3}\VANthebibliography}

\usepackage{graphicx}	
\usepackage{amsmath}	
\usepackage{amssymb}	
\usepackage{tabularx}
\usepackage{comment}
\usepackage{float}
\usepackage{xcolor}
\usepackage{orcidlink}
\usepackage{lineno}
\usepackage{soul}
\usepackage{cancel}
\usepackage[linesnumbered,ruled]{algorithm2e}
\usepackage{newtxtext,newtxmath}

\newcommand{\fref}{f_{\rm ref}}
\newcommand{\ARed}{A_{\rm sn}}
\newcommand{\gammaRed}{\gamma_{\rm sn}}
\newcommand{\ARedI}{A_{{\rm sn},I}}

\newcommand{\gammaRedI}{\gamma_{{\rm sn},I}}

\newcommand{\ADM}{A_{\rm DM}}
\newcommand{\ADMI}{A_{{\rm DM},I}}

\newcommand{\gammaDM}{\gamma_{\rm DM}}
\newcommand{\gammaDMI}{\gamma_{{\rm DM},I}}

\newcommand{\Pg}{P_{\rm g}}
\newcommand{\PI}{P_I}
\newcommand{\PJ}{P_J}
\newcommand{\PsnI}{P_{{\rm sn},I}}
\newcommand{\PdmI}{P_{{\rm DM},I}}
\newcommand{\chiIJhd}{\chi_{IJ}}
\newcommand{\chiIJdip}{\chi_{IJ}^{\rm dip}}

\newcommand{\NMeerTimePSRs}{88}
\newcommand{\TwoWkObsTimeHrs}{11.7}
\newcommand{\TwoWkObsTimeS}{42\,200}

\newcommand{\instBham}{Institute for Gravitational Wave Astronomy \& School of Physics and Astronomy, University of Birmingham, Birmingham, B15 2TT, UK}
\newcommand{\instSwin}{OzGrav-Swinburne, Centre for Astrophysics and Supercomputing, Swinburne University of Technology, PO Box 218, Hawthorn, Victoria 3122, Australia}
\newcommand{\instMelb}{OzGrav-Melbourne, School of Physics, University of Melbourne, Parkville, Victoria 3010, Australia}
\newcommand{\instOzGr}{ARC Centre of Excellence for Gravitational Wave Discovery (OzGrav), Mail H29, Swinburne University of Technology, PO Box 218, \\Hawthorn, Victoria 3122, Australia}
\newcommand{\instHull}{Centre of Foundation Year Studies, University of Hull, Cottingham Road, Hull HU6 7RX, UK}
\newcommand{\instMaxP}{Max-Planck-Institut f\"{u}r Radioastronomie, Auf dem H\"{u}gel 69, 53121, Bonn, Germany}
\newcommand{\instJodr}{Jodrell Bank Centre for Astrophysics, Department of Physics and Astronomy, University of Manchester, Manchester M13 9PL, UK}
\newcommand{\instAstr}{ASTRON, Netherlands Institute for Radio Astronomy, Oude Hoogeveensedijk 4, 7991 PD Dwingeloo, The Netherlands}
\newcommand{\instAmst}{Anton Pannekoek Institute for Astronomy, University of Amsterdam, Science Park 904, 1098 XH Amsterdam, The Netherlands}
\newcommand{\instINAF}{INAF-Osservatorio Astronomico di Cagliari, via della Scienza 5, 09047 Selargius (CA), Italy}
\newcommand{\instUoCT}{High Energy Physics, Cosmology and  Astrophysics Theory (HEPCAT) Group, Department of Mathematics and Applied Mathematics,\\ University of Cape Town, Rondebosch 7701, South Africa}
\newcommand{\instVand}{Department of Physics and Astronomy, Vanderbilt University, 2301 Vanderbilt Place, Nashville, TN 37235, USA}

\newcommand{\noBham}{1}
\newcommand{\noSwin}{2}
\newcommand{\noMelb}{3}
\newcommand{\noOzGr}{4}
\newcommand{\noINAF}{5}
\newcommand{\noUoCT}{6}
\newcommand{\noHull}{7}
\newcommand{\noMaxP}{8}
\newcommand{\noJodr}{9}
\newcommand{\noVand}{10}
\newcommand{\noAstr}{11}
\newcommand{\noAmst}{12}

\title[A simple optimisation for the MPTA]{A simple optimisation for the MeerKAT Pulsar Timing Array}

\author[H.~Middleton et al.]{Hannah~Middleton\orcidlink{0000-0001-5532-3622}$^{\noBham,\noSwin,\noMelb}$, \thanks{E-mail: h.r.middleton@bham.ac.uk (HM)}
Ryan~M.~Shannon\orcidlink{0000-0002-7285-6348}$^{\noSwin,\noOzGr}$, 
Matthew~Bailes\orcidlink{0000-0003-3294-3081}$^{\noSwin,\noOzGr}$, 
Andrew~D.~Cameron\orcidlink{0000-0002-2037-4216},$^{\noSwin,\noOzGr}$, 
\newauthor
Alessandro~Corongiu\orcidlink{0000-0002-5924-3141}$^{\noINAF}$, 
Marisa~Geyer\orcidlink{0000-0002-2822-1919}$^{\noUoCT}$, 
Max~Jones\orcidlink{0009-0005-9124-1348}$^{\noHull}$, 
Michael~Kramer\orcidlink{0000-0002-4175-2271}$^{\noMaxP,\noJodr}$, 
Matthew~T.~Miles\orcidlink{0000-0002-5455-3474}$^{\noVand,\noOzGr}$, 
\newauthor
Aditya~Parthasarathy\orcidlink{0000-0002-4140-5616}$^{\noAstr,\noAmst,\noMaxP}$, 
Andrea~Possenti\orcidlink{0000-0001-5902-3731}$^{\noINAF}$, 
Daniel~J.~Reardon\orcidlink{0000-0002-2035-4688}$^{\noSwin,\noOzGr}$
\\
$^{\noBham}$~\instBham\\
$^{\noSwin}$~\instSwin\\
$^{\noMelb}$~\instMelb\\
$^{\noOzGr}$~\instOzGr\\
$^{\noINAF}$~\instINAF\\
$^{\noUoCT}$~\instUoCT\\
$^{\noHull}$~\instHull\\
$^{\noMaxP}$~\instMaxP\\
$^{\noJodr}$~\instJodr\\
$^{\noVand}$~\instVand\\
$^{\noAstr}$~\instAstr\\
$^{\noAmst}$~\instAmst\\
}

\date{Accepted 2025 April 30. Received 2025 April 30; in original form 2024 June 20}

\pubyear{2025}

\begin{document}
\label{firstpage}
\pagerange{\pageref{firstpage}--\pageref{lastpage}}

\maketitle

\begin{abstract}
The goal of the MeerKAT radio telescope's pulsar timing array programme (MPTA) is the detection of gravitational waves (GWs) of nanohertz frequencies.
Evidence for such a signal was recently announced by the MPTA and several other pulsar timing array (PTA) consortia.
Given an array of pulsars and an observation strategy, we consider whether small adjustments to the observing schedule can provide gains in signal-to-noise ratio (S/N) for a stochastic GW background signal produced by a population of massive black hole binaries. 
Our approach uses a greedy algorithm to reallocate available integration time between pulsars in the array. 
The overall time dedicated to MPTA observing is kept constant so that there is only minimal disruption to the current observation strategy.
We assume a GW signal consistent with those reported.
For the sake of demonstrating our method, we also make several simplifying assumptions about the noise properties of the pulsars in the MPTA.
Given these assumptions, we find that small adjustments to the observing schedule can provide an increased S/N by $\approx 20\%$ for a $10\,{\rm yr}$ PTA lifespan.
\end{abstract}




\section{Introduction}
\label{sec:introduction}

Low-frequency gravitational wave (GW) detection is the objective of the global pulsar timing array (PTA) efforts. 
Decades-long PTA experiments are monitoring arrays of milli-second pulsars, searching for deviations in their pulse arrival times due to GWs~\citep{Sazhin:1978, Detweiler:1979, FosterBacker:1990}.
A prime target for GW searches with PTAs is the stochastic GW background (SGWB) at nanohertz frequencies, supposed to be predominantly produced by a cosmological population of merging massive black hole binaries (MBHBs)~\citep[see, for example,][]{BurkeSpolaorEtAl:2019,Sesana:2013, RajagopalRomani:1995}.

The MeerKAT radio telescope is a $64$-antenna interferometer located in the Karoo region in the Northern Cape of South Africa.
MeerTime is a South African Radio Astronomy Observatory (SARAO) Large Survey Project, which is centred around the MeerKAT pulsar timing program. 
For an overview of MeerTime, see~\cite{MeerTimeBailesEtAl:2016, MeerTimeBailesEtAl:2020}. 
MeerTime pulsar science includes the study of relativistic pulsar binaries~\citep{MeerTimeRelBin:2021}, discovery and timing of globular cluster pulsars~\citep{MeerTimeGlobularClusterPSRs:2021}, pulsar phenomena studies with the thousand pulsar array~\citep{MeerTime1000PSR:2020,MeerTime1000PSR:2021}, and the MeerKAT PTA (MPTA)~\citep{MPTANoiseMilesEtAl:2025, MPTAGWMilesEtAl:2025, MPTAMapsGrunthalEtAl:2025, MilesEtAl:2023,SpiewakEtAl:2022} which is the focus of this work. 

Recent PTA results indicate evidence for a GW detection of $2\sigma$--$4\sigma$ significance from the MPTA~\citep{MPTAGWMilesEtAl:2025}, the Parkes PTA~\citep[PPTA;][]{PPTAGWReardonEtAl:2023}, the European and Indian PTAs combined~\citep[EPTA+InPTA;][]{ETPAInPTAGWBAntoniadisEtAl:2023}, the North American Nanohertz Observatory for GWs~\citep[NANOGrav;][]{NANOGravGWBAgazieEtAl:2023}, and the Chinese PTA~\citep[CPTA;][]{CPTAGWBXuEtAl:2023}.
Further study will be required to confirm the nature of this signal~\citep[see, for example,][]{ZicEtAl:2022,GoncharovEtAlOnPPTA:2021}, including analysis of the combined PTA datasets as part of the full International PTA~\citep[IPTA;][]{IPTADR2:2022, IPTA:2016, ChenWuHuangIPTADR2:2021, AgaziEtlAl:2024}. 
Assuming a MBHB origin, the observation of a GW signal of this type provides insights into the lives and population of MBHBs and the formation and evolution of galaxies~\cite[e.g.][]{EPTAInPTAInterp:2023, NANOGravMBHBInterp:2023, SteinleEtAl:2023, BarausseEtAl:2023, MiddEtAl:2021, ChenSesanaConselice:2019}. 
As PTAs share common MBHB sources with the future space-based Laser Interferometer Space Antenna~\citep[LISA;][]{LISA:2024}, there will also be the opportunity for complementary observations~\citep{SteinleEtAl:2023} upon LISA's launch in the mid-2030s.
Other possible GW origins for the signal in PTA data include processes occurring in the early Universe and/or involving Dark Matter~\citep{EPTAInPTAInterp:2023, NANOGravNewPhysics:2023}.
For the purposes of this work, we focus on the MBHB population. 

Predicting the anticipated amplitude of the SGWB is an area of active research~\citep[see, for example][]{JaffeBacker:2003, SesanaExpectedGWB:2013, KelleyEtAl:2017, SykesEtAl:2022}.
Equally, modelling the anticipated sensitivity of a PTA is an important task to inform observation scheduling~\citep{ChristyEtAl:2014,Lam:2018} and allow time-to-detection forecasting~\citep{TaylorEtAl:2016,VigelandSiemens:2016,KelleyEtAlTimeToDetection:2017, PolEtAl:2021,BecsyEtAl:2022}.
As an example, it has been shown that increasing the number of \emph{good} pulsars in the array (i.e. those providing low enough root-mean-square uncertainty in their residuals over sufficiently long timescales) provides good returns on sensitivity improvements for SGWB searches~\citep[e.g.][]{SiemensEtAl:2013}.
MeerKAT's high sensitivity~\citep{MeerTimeBailesEtAl:2020} means shorter observation times for the same root-mean-square uncertainty for a pulsar. 
This allows for more pulsars to be timed, making MeerKAT an excellent facility for PTA observations, however we also note that shorter integration times can be more impacted by jitter noise~\cite[see, e.g.][]{CordesShannon:2010, CordesDowns:1985, ParthasarathyEtAl:2021}.

In this work we consider whether small adjustments to the current MPTA observing strategy can provide improvements in signal-to-noise ratio (S/N) for a SGWB signal. 
Numerous previous works have also considered PTA optimisation, including building scaling relations for time-to-detection estimation~\citep{SiemensEtAl:2013}, schedule optimisation for single and multiple telescopes~\citep{LeeEtAl:2012, ChristyEtAl:2014, Lam:2018, LamEtAl:2018}, and pulsar selection and coverage~\citep{SperiEtAl:2023, Roebber:2019}.   
Here, our starting point is the existing MPTA observation schedule and our aim is to investigate whether minor changes can improve S/N for the PTA. 
We take a greedy algorithm approach~\citep[e.g.][]{DechterDechter:1998} and only make changes to the time allocated to each pulsar, \emph{i.e.}, we do not consider whether a pulsar should be discarded from the PTA. 

In Section~\ref{sec:snr} we describe how the PTA S/N is computed.
In Section~\ref{sec:data} we describe the MPTA data used in this work and the anticipated sensitivity for the current observation strategy. 
We note that, while MPTA pulsars are the focus of this work, the methods might be extended to the full IPTA observation strategy. 
We investigate small adjustments to the observing schedule in Section~\ref{sec:timeshuffle} and we present the results in Section~\ref{sec:results}.
Finally, in Section~\ref{sec:twentyPercent} we investigate the S/N gains possible from an additional allocation in observing time and present the conclusions in Section~\ref{sec:conclusions}.

\section{PTA average signal-to-noise (S/N) ratio}
\label{sec:snr}

We compute the average S/N for a given PTA and a predicted SGWB signal following the method in~\cite{SiemensEtAl:2013}. 
The power spectrum for an SGWB is given by
\begin{equation}
\Pg (f) = \frac{A^2}{24 \pi^2} \left(\frac{f}{\fref}\right)^{2\alpha} f^{-3}\,,
\label{eqn:Pgw}
\end{equation}
where $f$ is the GW frequency, $\fref$ is a reference frequency chosen as $1\,{\rm yr}^{-1}$, $A$ is the amplitude of the SGWB at $f=\fref$, and $\alpha$ is the slope of the spectrum. 
Throughout this work we assume that $\alpha=-2/3$, consistent with a stochastic background from circular MBHBs, and that $A = 2 \times 10^{-15}$, consistent with recent observational evidence for a GW detection ~\citep{MPTAGWMilesEtAl:2025, PPTAGWReardonEtAl:2023, ETPAInPTAGWBAntoniadisEtAl:2023, NANOGravGWBAgazieEtAl:2023, CPTAGWBXuEtAl:2023, IPTADR2:2022,GoncharovEtAlOnPPTA:2021, ChenEtAlEPTA:2021, NANOGrav12.5GWB:2020}.

The average S/N for a PTA with total timespan $T$ is given by~\citep{SiemensEtAl:2013}
\begin{equation}
\left< \rho \right> = \left( 2 T \sum_{IJ} \chiIJhd^2 
                             \int_{f_L}^{f_H} {\rm d}f  
                             \frac{\Pg^2 (f)}{\PI(f)\PJ(f)}
                            \right)^{1/2}\,,
\label{eqn:snrPIPJPg}
\end{equation}
where $\sum_{IJ}$ is the sum over all pulsar pairs ($I \neq J$) and $P_I$ is the power spectrum of the $I{\rm th}$ pulsar's timing residuals. 
We make the simplifying assumption that $T$ is the same for all pulsars, \emph{i.e.} all pulsars are part of the array from the beginning. 
The integration limits are given by $f_L=1/T$ and $f_H=\kappa/2$ where $\kappa$ is the cadence of the observations, which we also assume to be the identical for all pulsars in the array. 
The Hellings and Downs coefficients $\chi_{IJ}$ describe the expected angular correlation produced by the SGWB signal in pairs of pulsars and is given by
\begin{eqnarray}
\chiIJhd &=& \frac{1}{2} 
           - \frac{1}{4} \left(\frac{1 - \cos(\zeta)}{2}\right) \nonumber\\
           && + \frac{3}{2} \left(\frac{1 - \cos(\zeta)}{2}\right) 
           \log \left( \frac{1 - \cos(\zeta)}{2} \right)\,,
\label{eqn:hd}
\end{eqnarray}
for pulsars $I$ and $J$ ($I \neq J$) separated by angle $\zeta$~\citep{HellingsDowns:1983}.

The overall power spectrum for a pulsar, $\PI (f)$, includes the contribution from the SGWB and several noise sources~\cite[see][for a review of pulsar noise]{WangNoiseReview:2015}.
For the purpose of this study, we include noise due to timing precision, pulse jitter, spin irregularities, and dispersion measure (DM) variations. 
The resulting $\PI (f)$ for each pulsar is given by
\begin{eqnarray}
\PI (f) &=& \Pg (f) ~+~ 2 \left( \sigma_I^2 + \sigma_{{\rm jit}, I}^2 \right) \Delta t \nonumber\\
        & &~+~\PsnI (f) ~+~ \PdmI (f) \,,
\label{eqn:PIGWWR}
\end{eqnarray}
where $\sigma_I$ is the timing precision for the $I{\rm th}$ pulsar and $\Delta t=1/\kappa$. 
We make the assumption of evenly sampled data. 
Each of the other terms in equation~(\ref{eqn:PIGWWR}) are described in turn below.

Jitter noise $\sigma_{\rm jit}$ is caused by stochastic changes in the individual pulse profile shape~\citep[e.g][]{CordesDowns:1985, ParthasarathyEtAl:2021}.
The noise is added in quadrature to the timing precision in equation~(\ref{eqn:PIGWWR}).
For simplicity, we do not consider ECORR noise~\cite[see, e.g.,][]{GoncharovEtAl:2021} in this work.

Spin noise, (also called red or timing noise) arises due to irregularities in the rotation rate of the pulsar~\cite[e.g.][]{ShannonCordes:2010}. 
The power spectral density of spin noise for the $I{\rm th}$ pulsar, $\PsnI (f)$, is given by~\cite[see, e.g.,][]{PPTAGWReardonEtAl:2023, GoncharovEtAl:2021}
\begin{equation}
\PsnI (f) = \frac{\ARedI^2}{12 \pi^2} 
            \left( \frac{f}{\fref} \right)^{-\gammaRedI} 
            {\rm yr}^3\,,
\end{equation}
where $\ARedI$ and $\gammaRedI$ are the amplitude and spectral index, respectively.

Finally, DM noise is a chromatic noise arising due to variable dispersion of the pulsar signal as it travels through the ionised interstellar medium~\cite[e.g.][]{Cordes:2002, TiburziEtAl:2021}. 
The power spectral density of DM noise for the $I{\rm th}$ pulsar, $\PdmI (f)$ is similarly modeled as a power law~\cite[see, e.g.,][]{PPTAGWReardonEtAl:2023}
\begin{equation}
\PdmI (f,\nu) = \frac{\ADMI^2}{12 \pi^2} 
            \left(\frac{\nu}{1,400\,{\rm MHz}}\right)^{-2}
            \left( \frac{f}{\fref} \right)^{-\gammaDMI} 
            {\rm yr}^3\,,
\end{equation}
where $\ADMI$ and $\gammaDMI$ are the amplitude and spectral index, respectively.
We use a reference observation frequency of $\nu=1,400\,{\rm MHz}$~\cite[as in][as the L-band receiver is used for MPTA analysis]{MPTANoiseMilesEtAl:2025}.

Throughout this analysis, we use equations~(\ref{eqn:snrPIPJPg}) and~(\ref{eqn:PIGWWR}) to compute the average PTA S/N for a given observation strategy, pulsar noise parameters (see Section~\ref{sec:data}), and assumed SGWB signal.

\section{Data}
\label{sec:data}
For the purpose of this study, we take a realisation of the MPTA consisting of $\NMeerTimePSRs$ pulsars, each of which has an estimated $\sigma_I$ and integration time (for a single pulsar observation) required to achieve that $\sigma_I$.
From the radiometer equation, the value of $\sigma_I$ scales with the instrument system temperature and with the inverse square root of both the integration time $T_{{\rm obs},I}$ and the receiver bandwidth~\citep[see, for example][]{LorimerKramer:2012, HandzoEtAl:2015}. 
We estimate $\sigma_I$ as
\begin{equation}
\sigma_I = \frac{\delta_I}{\sqrt{T_{{\rm obs},I}}}\,,
\label{eqn:TobsToSig}
\end{equation}
where $\delta_I$ is a constant of proportionality which is different for each pulsar. 
Throughout this work we make the simplifying assumption that $\delta_I$ is a constant for each pulsar throughout the PTA lifespan.
This simplification also relies on the assumption that identical detection apparatus is used for each observation throughout the PTA lifespan. 

Equation~(\ref{eqn:PIGWWR}) shows that the average PTA S/N depends on noise contributions from spin irregularities, DM variations, and pulse jitter of each pulsar in the array.
The first MPTA data release~\citep[DR1][]{MilesEtAl:2023} provides estimated spin noise  parameters for several of the pulsars considered here. 
MeerKAT $\ARed$ and $\gammaRed$ measurements exist for $4$ of the $\NMeerTimePSRs$ pulsars used here\footnote{We note that this analysis was performed before the release of the MPTA GW search results, so values available at that time are used in this demonstration of the method and the task of refining the optimisation with new values is left to future study.}. 
We also use estimates provided by noise analyses from other PTAs where available: $26$ pulsars from the PPTA~\citep{PPTANoiseReardonEtAl:2023}, $5$ pulsars from NANOGrav~\citep{NANOGravNoise:2023}, and $2$ pulsars from EPTA+InPTA~\citep{EPTANoise:2023}. 
Where multiple PTAs have measured the spin noise for the same pulsar, we choose the measurement with the lowest $A_{\rm sn}$ value. 
This is an optimistic assumption in order to avoid unduly disfavouring a pulsar in the time-swap analysis (see Section~\ref{sec:timeshuffle}).

Similarly, the MPTA DR1 provides $\ADM$ and $\gammaDM$ estimates for $31$ of the pulsars considered here. 
Again, we also use DM noise estimates from the PPTA ($15$ pulsars) and EPTA+InPTA ($4$ pulsars), where available~\citep{PPTANoiseReardonEtAl:2023,EPTANoise:2023}\footnote{We do not use NANOGrav DM values in this work as they use a different formalism for DM measurements, see e.g.~\cite{NANOGrav:2015}}. 

We use jitter noise estimates for $13$ of the pulsars taken from~\cite{ParthasarathyEtAl:2021}.
In cases where the jitter noise values are upper limits, we treat them in the same way as those pulsars without published noise values for simplicity (see below). 

Of the $\NMeerTimePSRs$ pulsars considered here, we therefore have spin noise estimates for $37$ pulsars, DM noise estimates for $50$ pulsars, and jitter noise estimates for $13$ pulsars. 
We use only the quoted noise measurements without errors in this work, \emph{i.e.}, we do not factor in the uncertainty in this analysis.

Some pulsars have no published noise parameters. 
In these cases, we use assumed noise values informed by the pulsars that do have published noise parameters. 
The choice is arbitrary. 
For pulsars without published $\gamma_{\rm sn}$ and $\gamma_{\rm DM}$ values, we take the median of the published values of $\gamma_{\rm sn}$ and $\gamma_{\rm DM}$, respectively.  
For pulsars without published $\ARed$, $\ADM$, and $\sigma_{\rm jit}$, we take the minimum of the published values of $\ARed$, $\ADM$, and $\sigma_{\rm jit}$, respectively. 
Assigning the minimum value for the $\ARed$, $\ADM$, and $\sigma_{\rm jit}$ parameters is an optimistic choice which avoids setting noise parameters unrealistically to zero, whilst not unduly disfavouring a pulsar by giving it a large noise value.

The MPTA observation strategy we start from targets a precision of $\sigma_I = 1\,{\rm \mu s}$ for every pulsar with a cadence of one observation every two weeks ($\kappa=26\,{\rm yr}^{-1})$.  
The total observation time used in each two week observing time is $\TwoWkObsTimeS\,{\rm s} \approx \TwoWkObsTimeHrs\,{\rm hrs}$, which is divided between the MPTA pulsars.

Assuming an SGWB signal with $A = 2\times 10^{-15}$ and $\alpha = -2/3$, we can estimate the predicted average S/N for the MPTA from equation~(\ref{eqn:snrPIPJPg}) after a timespan of $T=10\,{\rm yrs}$.
Using the noise properties assumed above, the resulting S/N at $10\,{\rm yrs}$ is $4.4$.
However, we note that the actual S/N values and the results of the analysis presented here are heavily dependent on the assumed noise parameters for each pulsar.
In the following sections, we therefore focus on the percentage gains in S/N as a result of our analysis and not the absolute value of the S/N.
As described above, we assign minimal values for those pulsars that have no current noise measurements. 
The reliability of these S/N computations will be improved as further noise measurements become available. 
In the following sections we focus on whether the average S/N can be improved by minor changes to the integration time allocated to each pulsar.

\section{The time swap method}
\label{sec:timeshuffle}

We take a greedy algorithm~\citep{DechterDechter:1998} approach to find S/N improvements to the MPTA.
A greedy algorithm is one which takes the locally optimal choice at each stage. 
The approach has its limitations in that always taking the locally optimal step can lead to missing the global optimal solution. 
However, it has a reasonably low computational cost and can still achieve S/N improvement. 
Here we describe the method and in Section~\ref{sec:results} we present our results.

The time-swap method we apply works by swapping a fraction of the time allocation between pulsars. 
We use the integration times from in the current observing strategy as a starting point (see Section~\ref{sec:data}).
The total observation time available to MPTA is kept constant ($\sum_{I} T_{{\rm obs},I} = \TwoWkObsTimeS\,{\rm s}$) and the session cadence is fixed at $\kappa=26\,{\rm yr}^{-1}$ (i.e., about every two weeks)~\citep{SpiewakEtAl:2022}, which means that only limited changes to the current observing schedule are possible. 
We also place the additional constraint of a minimum integration time per pulsar of $T_{{\rm obs},I} \geq 256\,{\rm s}$ so that no pulsar is removed entirely from the array and to reduce jitter noise to some extent.
Note that some particularly bright pulsars need only $256\,{\rm s}$ of integration time to obtain the targeted timing precision of $\sigma_I = 1\,{\rm \mu s}$ with the MTPA.
Integration time is swapped between pulsars to find an S/N improvement. 
At each step, a fraction $h$ of integration time is subtracted from a pulsar ($I$) and reallocated to another pulsar ($J$) so that 
\begin{eqnarray}
T_{{\rm obs},I} &\rightarrow& T_{{\rm obs},I} - h T_{{\rm obs},I}\,, \nonumber\\
T_{{\rm obs},J} &\rightarrow& T_{{\rm obs},J} + h T_{{\rm obs},I}\,, \nonumber
\end{eqnarray}
where we choose $h=1/8$. 
Every possible combination of time swaps between pulsar pairs is trialled.
The swap which gives the biggest S/N increase at $10\,{\rm yrs}$ is selected and the list of $T_{{\rm obs}}$ is updated accordingly.
The new time allocations become the starting point for the next time swap. 
The process is repeated until the algorithm cannot find further gains in S/N. 
A simplified version of the method is shown as pseudocode in algorithm~\ref{alg:timeswap}.

\begin{algorithm}
  \underline{time swap method}\\  
  \While {improvement==True}
    {
    \For {ipsr {\rm \textbf{in}} pulsarList} {
      \For {jpsr {\rm \textbf{in}} pulsarList (ipsr!=jpsr)} {
        reassign fraction of integration time (ipsr $\rightarrow$ jpsr) \\
        compute S/N given new integration times \\
        \If {this S/N is largest so far} {
          save result\\
          move on to next psr 
          }
      }
    }
    \eIf {improvement was made} {
      update integration times with best result\\
      repeat while loop 
      }
      {
      save integration times \\
      improvement == False (exits while loop)
      }   
    }
    \caption{Pseudocode for the time swap method used here.}
    \label{alg:timeswap}
\end{algorithm}

\begin{figure}
\includegraphics[width=0.5\textwidth]{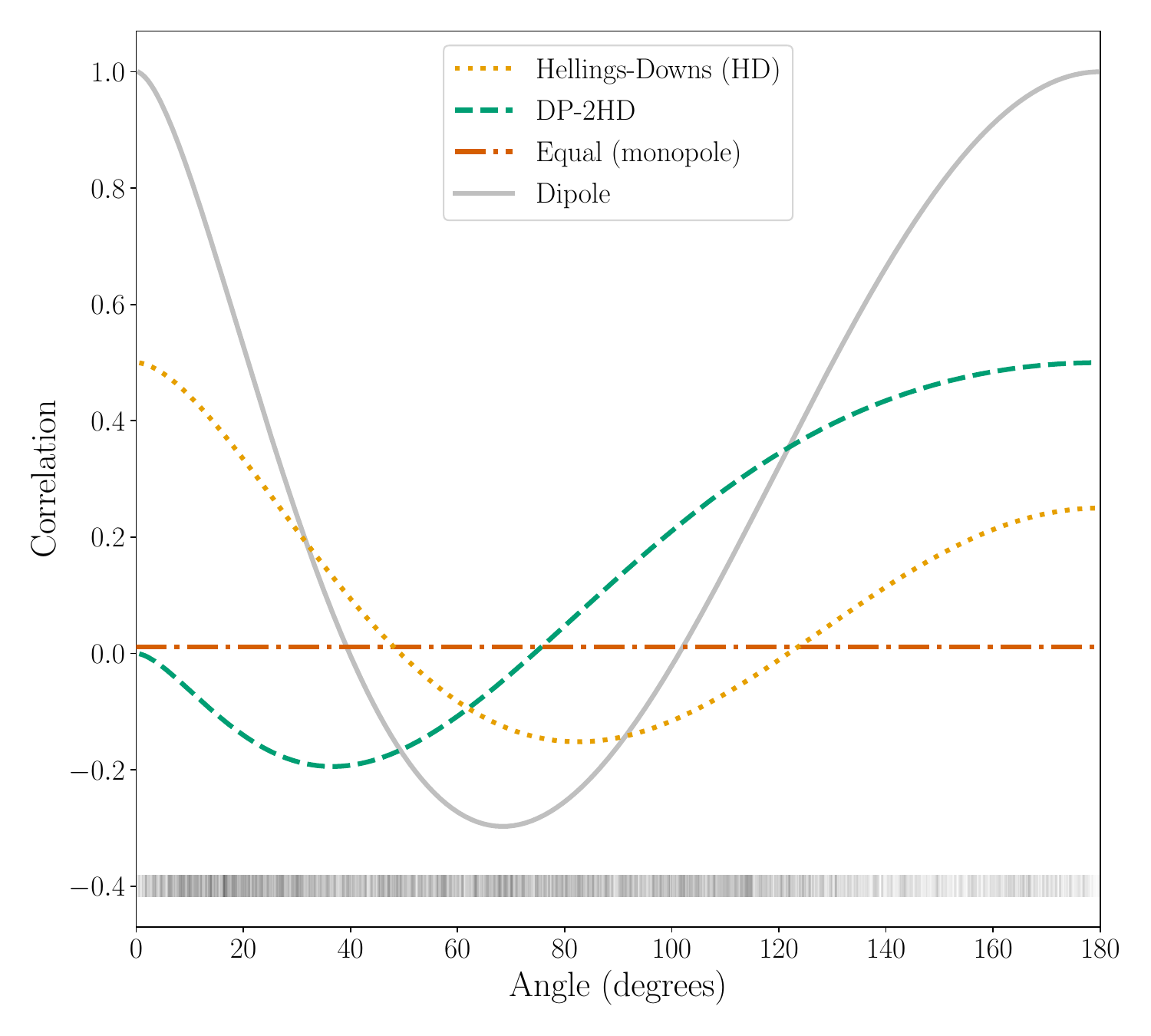}
\caption{\label{fig:correlations}
Correlation functions. 
The orange-dotted line shows the Hellings-Downs curve, equation~(\ref{eqn:hd}). 
The grey-solid line shows a dipole correlation, equation~(\ref{eqn:dip}).
The green-dash line shows the function constructed in equation~(\ref{eqn:chimod}) to favour pulsars at larger angular separations.
The red-dot-dash line shows an equal (monopole) correlation, which assigns equal values to all pulsar pairs independent of angular separation (this is arbitrarily set to one over the number of pulsars in our analysis.)
The grey markers at the bottom of the plot represent the angular coverage of the pulsar pairs used in this analysis as a one-dimensional density plot (the darker shades represent a greater number of pulsar pairs).
}
\end{figure}

The shape of the Hellings-Downs function [equation~(\ref{eqn:hd})] means that pulsars pairs with small angular separations have higher contributions to the S/N in equation~(\ref{eqn:snrPIPJPg}). 
The Hellings-Downs curve is shown by the orange-dotted line in Fig.~\ref{fig:correlations}.
Looking ahead to Section~\ref{sec:results}, we find that pulsars towards the Galactic centre are favoured in the time swap analysis due to the large number of pulsar pairs with small angular separations.
It is, however, important for a PTA to have a good coverage of angular separations so that a Hellings-Downs GW signature can be confirmed. 
We therefore construct two alternative correlation functions to reduce bias towards Galactic centre pulsars and repeat the analysis.
The first is a dipole $\chiIJdip$ minus a Hellings-Downs curve (labelled `DP-2HD') given by
\begin{equation}
\chi_{IJ}^{\rm mod} = \chiIJdip - 2 \chiIJhd\,,
\label{eqn:chimod}
\end{equation}
where $\chiIJdip$ is given by~\citep{AnholmEtAl:2009},
\begin{equation}
\chiIJdip = -\frac{3}{2}
            \left[ \cos \zeta - \frac{4}{3} - 4 \tan^2 \frac{\zeta}{2} 
                   \ln \left( \sin \frac{\zeta}{2} \right) \right]\,.
\label{eqn:dip}
\end{equation}
Equation~\ref{eqn:chimod} is shown by the green-dash line in Fig.~\ref{fig:correlations}. 
In this case, pulsar pairs at wider separations are favoured.
The other alternative correlation function we consider sets all values of $\chiIJhd$ to an equal value \emph{(i.e.}, a monopole), removing the sky position bias completely (which is represented as the red-dot-dash horizontal line at an arbitrary constant value in Fig~\ref{fig:correlations}). 

The alternative correlation functions described are used to replace $\chi_{IJ}$ in equation~(\ref{eqn:hd}), providing a proxy for S/N which is less biased toward pulsars in the Galactic centre. 
We note that this proxy for the S/N is only used within the time-swap method to perform the S/N improvement checks (\emph{i.e.,} lines $6$--$10$ in algorithm~\ref{alg:timeswap}). 
When assessing the overall S/N gain of the time-swap analysis in the following sections, we use the resulting integration times to recompute the S/N using the Hellings-Downs correlation function in order to provide a fair comparison between the potential gains in S/N for a SGWB signal.

\section{Results}
\label{sec:results}

\begin{figure}
\includegraphics[width=0.5\textwidth]{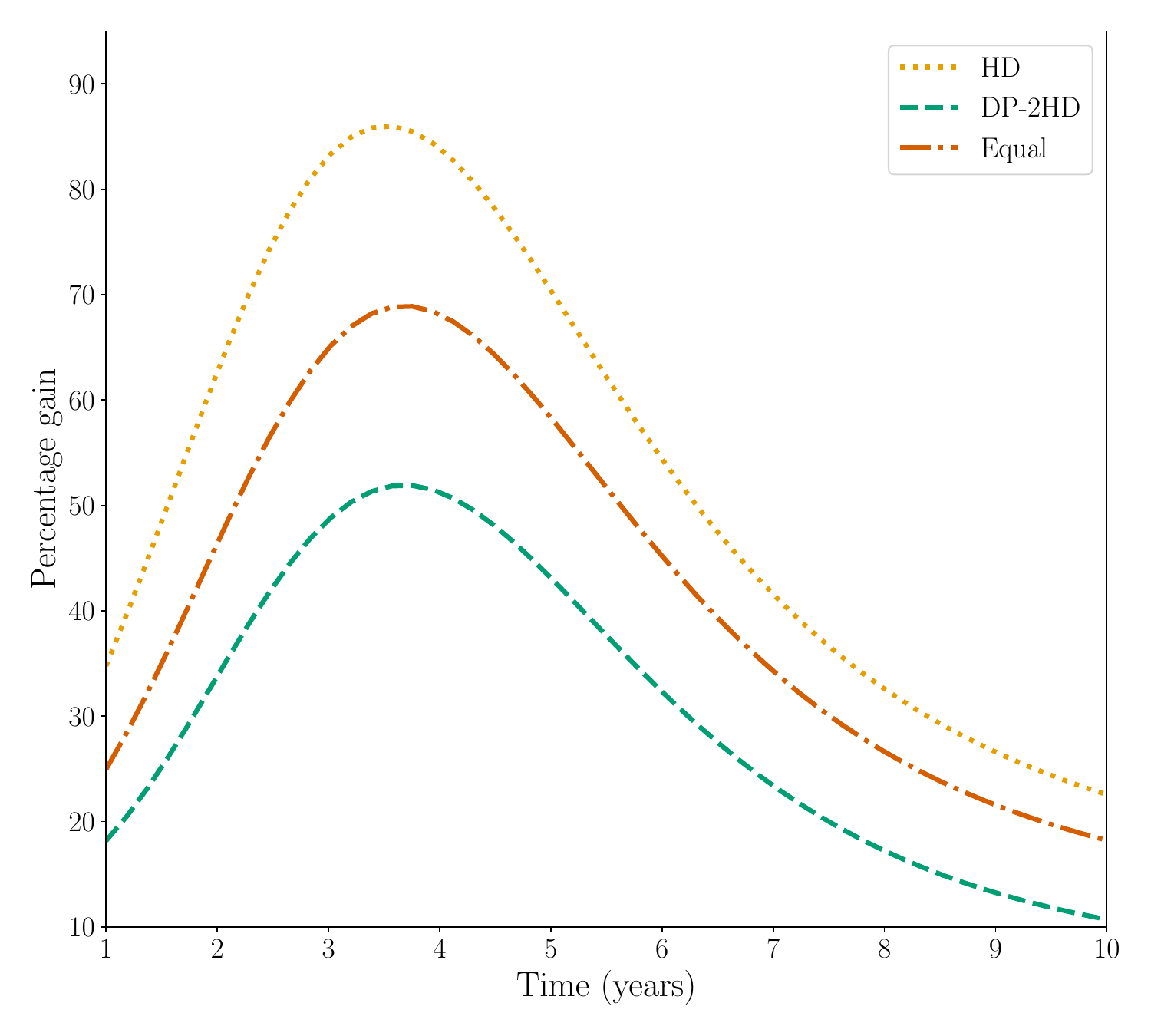}
\caption{\label{fig:snrCompare}
Results of the time-reallocation analysis. 
The percentage S/N gain over the original observing strategy is shown over time up to an observing time of $10\,{\rm yr}$.
Three time reallocation results are shown which each assume a different function for $\chiIJhd$ in the time-swap analysis.
`HD' is Hellings-Downs, `DP-2HD' is the difference between dipole and Hellings-Downs, and `Equal' is equal value for all pulsar pairs (\emph{i.e.,} a monopole). 
Note that the alternative correlation functions are used only within the time-swap analysis. 
We use the resulting integration times to recompute the S/N over time for a SGWB (using the Hellings-Downs correlation function) in order to make a fair comparison between the time-swap results and the original strategy.
}
\end{figure}

\begin{figure*}
HD\\
\includegraphics[trim={0cm 4cm 0cm 3.5cm},clip,width=0.7\textwidth]{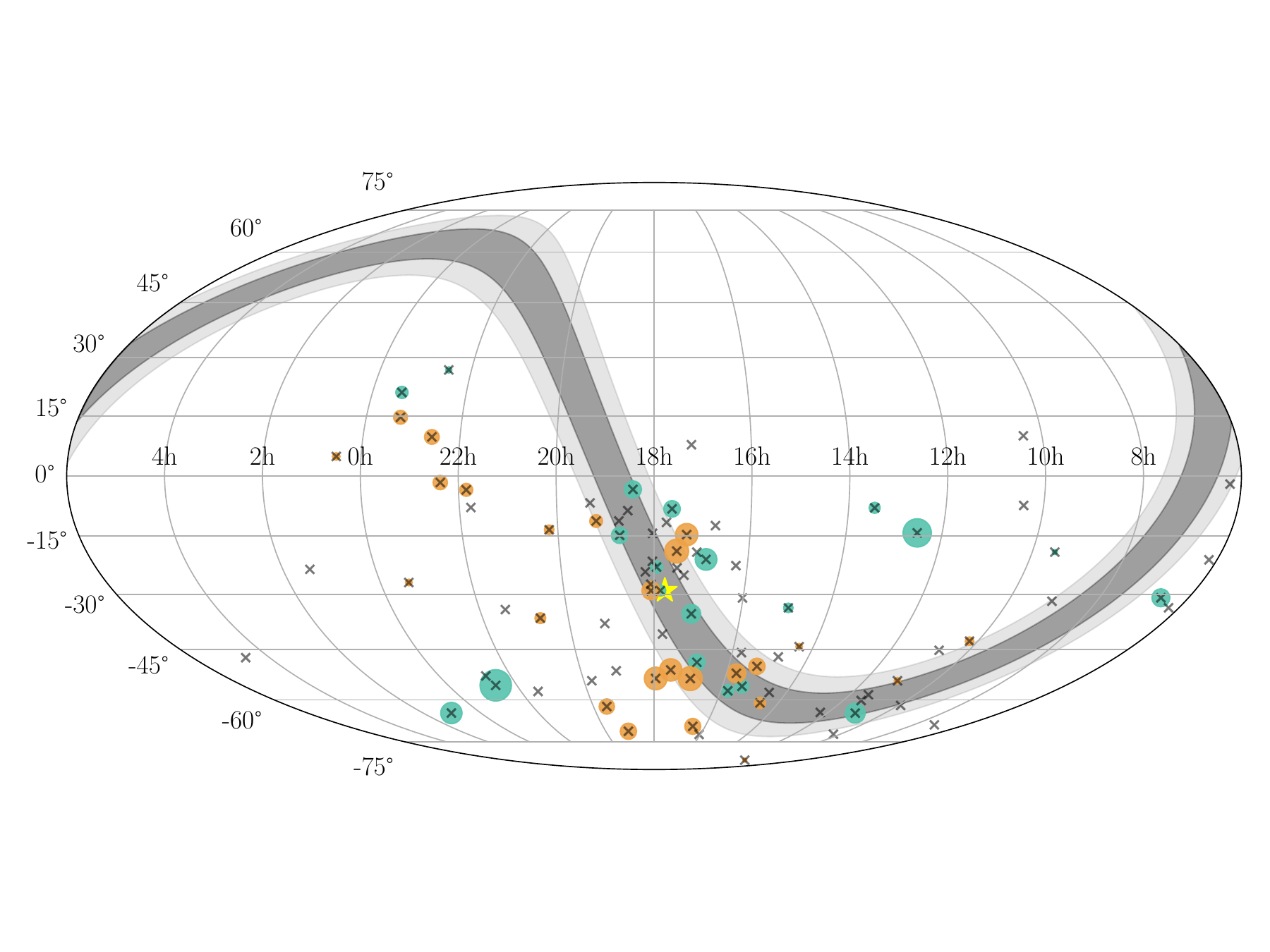}\\\vspace{2.5em}
DP-2HD\\
\includegraphics[trim={0cm 4cm 0cm 3.5cm},clip,width=0.7\textwidth]{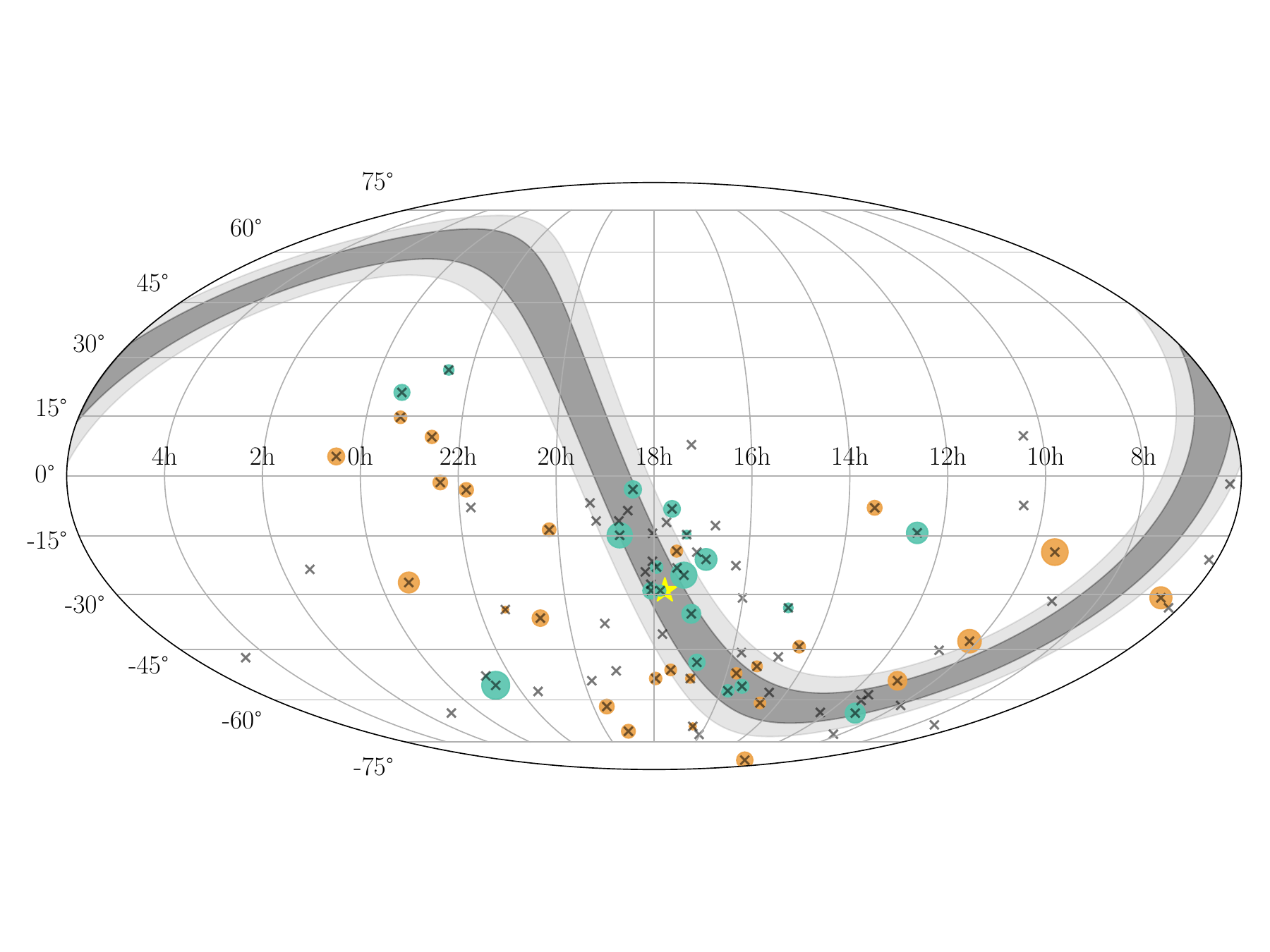}\\\vspace{2.5em}
Equal\\
\includegraphics[trim={0cm 4cm 0cm 3.5cm},clip,width=0.7\textwidth]{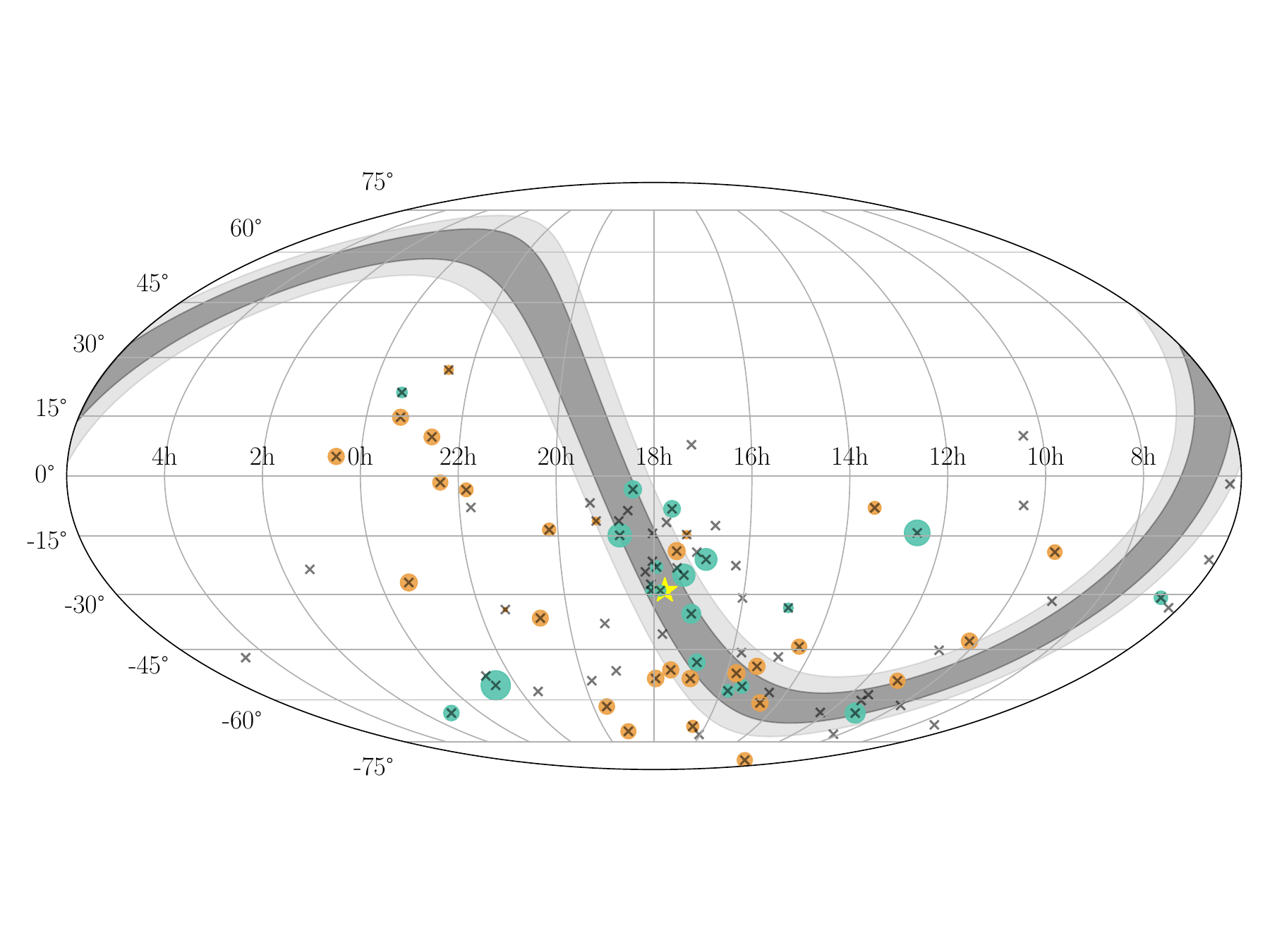}
\caption{\label{fig:skymaps}
Sky map showing changes to the integration times for each pulsar after the time-swap method. 
Each cross marker indicates the position of a pulsar. 
Orange and green circles indicate pulsars that have gained and lost integration time, respectively. 
The size of the circle is proportional to the change in integration time. 
A marker with no green or orange circle indicates a pulsar whose integration time was unchanged by the analysis (or the change was too small to be plotted).
The yellow star and grey band indicate the Galactic centre and Galactic plane, respectively. 
Top (HD): results using the Hellings-Downs correlation function in the time swap show that pulsars in the Galactic centre are favoured, gaining more time. 
Centre (DP-2HD): results using a correlation function that favours wide angular separations between pulsars [equation~(\ref{eqn:chimod})]. 
Pulsars in the Galactic centre are disfavoured in this case.
Bottom (Equal): results using an equal correlation function.
Pulsars in the Galactic centre are no longer favoured or disfavoured due the the small angular separation of pairs.
}
\end{figure*}

Results of the time swap analysis are shown in Fig.~\ref{fig:snrCompare} in which the percentage average S/N gain over the original observing strategy over a $10\,{\rm yr}$ PTA lifespan is shown. 
The results of time swapping with three different correlation functions are shown: Hellings-Downs (orange-dot), dipole minus Hellings-Downs (green-dash), and Equal (red-dot-dash).
We again note that the alternative correlation functions are only used during the time-swap calculation. 
Regardless of correlation function used, the output of the method is a new list of pulsar integration times, from which we compute the S/N over $10\,{\rm yrs}$ for a SGWB using equations~(\ref{eqn:snrPIPJPg}) and~(\ref{eqn:hd}).
This provides a fair comparison between the three time-swap analyses performed here and the original observing strategy.

The biggest S/N gain is achieved using the Hellings-Downs function for the time-swap analysis. 
The S/N gain at $T=10\,{\rm yrs}$ is $\approx 22\%$ over the original observation strategy.
However, as discussed in Section~\ref{sec:timeshuffle}, this comes at the cost of reducing the integration time for pulsar pairs with high angular separations, which could make confirming a Hellings-Downs correlation more challenging. 

This effect of using a Hellings-Downs correlation in the time swap is clear from the sky maps shown in Fig.~\ref{fig:skymaps}. 
The position of each pulsar is shown by a cross marker. 
Changes in integration time (compared to the original time allocations) are shown by coloured circles. 
Pulsars that gain integration time are shown in orange and those that lose it in green. 
The size of the circle is proportional to the amount of time gained or lost. 
The top panel in Fig.~\ref{fig:skymaps} shows the result of using the Hellings-Downs correlation in the time swap. 
The Galactic centre is indicated by the yellow-star marker and the galactic plane by the grey band. 
The general trend we see is that pulsars close to the Galactic centre in projection gain time and those away from it lose time. 
The other two correlation function are shown in the centre and bottom panel in Fig.~\ref{fig:skymaps}. 
In both the DP-2HD and Equal cases, we see less of a bias towards pulsars in the Galactic centre than that seen in the top panel, however we also note that the DP-2HD option may be too strongly biased against pulsars in the Galactic centre.
We therefore suggest that the Equal option provides the most unbiased approach in terms of pulsar sky position. 

Returning to Fig.~\ref{fig:snrCompare}, we see that the second-largest S/N gain is from using the Equal correlation function in the time swap, giving an increase in S/N of $\approx 18\%$ after $10\,{\rm yrs}$. 
Finally, using the DP-2HD correlation function provides an increase of $\approx 10\%$ after $10\,{\rm yrs}$. 
Although using a Helling-Downs correlation function provides the largest S/N increase in Fig.~\ref{fig:snrCompare}, we recommend replacing it with an Equal correlation function for the time-swap calculation to prevent preference for pulsar pairs with small angular separations.

Figure~\ref{fig:snrCompare} also shows that the S/N gains resulting from time reallocation could be greater than $50\%$ in the short term, i.e., $2.5$ to $5\,{\rm yr}$.
However, the shorter term gains for all reallocation methods level out towards the end of the $10\,{\rm yr}$ period considered here. 
We do not find an intuitive explanation for the higher short-term gain and leave this as a topic for future study.

\begin{figure*}
\includegraphics[width=0.49\textwidth]{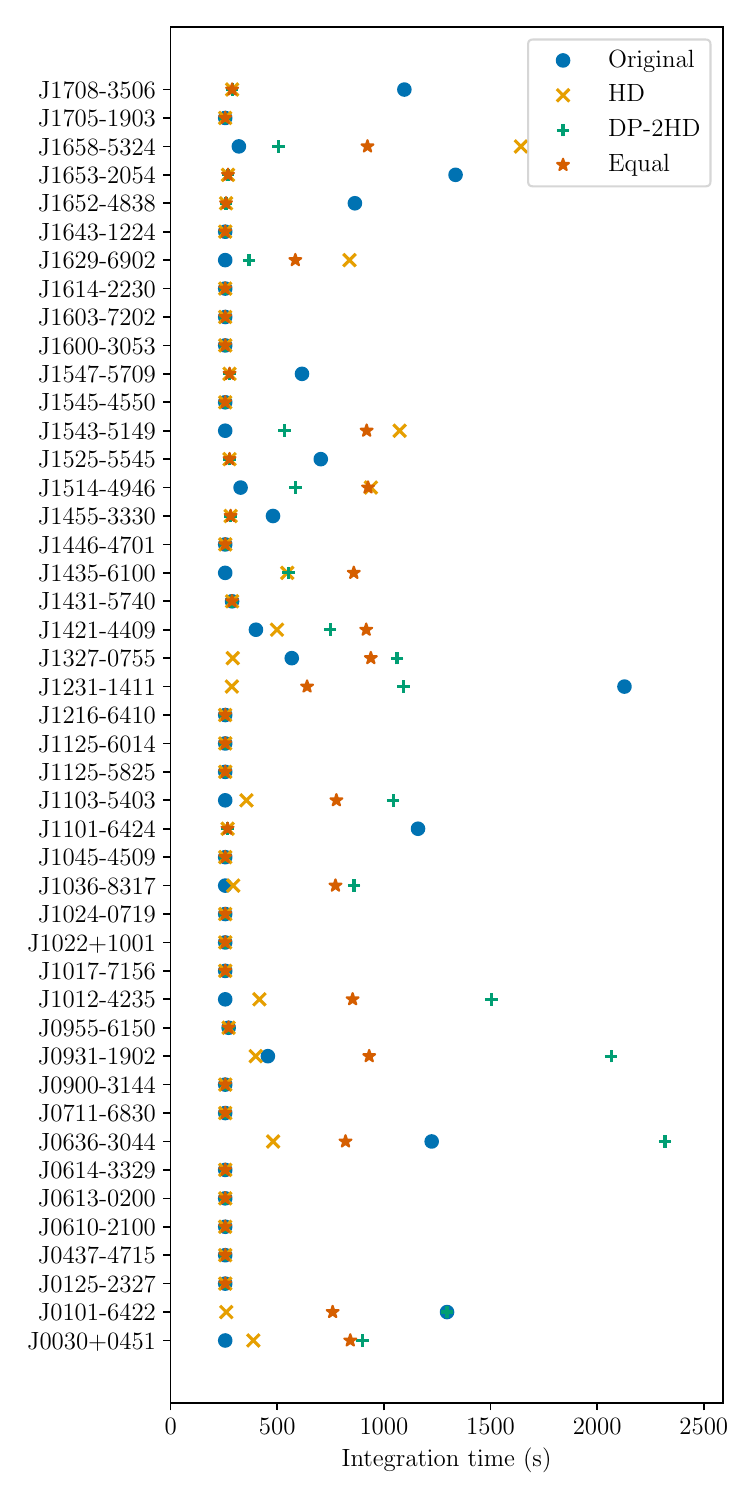}
\includegraphics[width=0.49\textwidth]{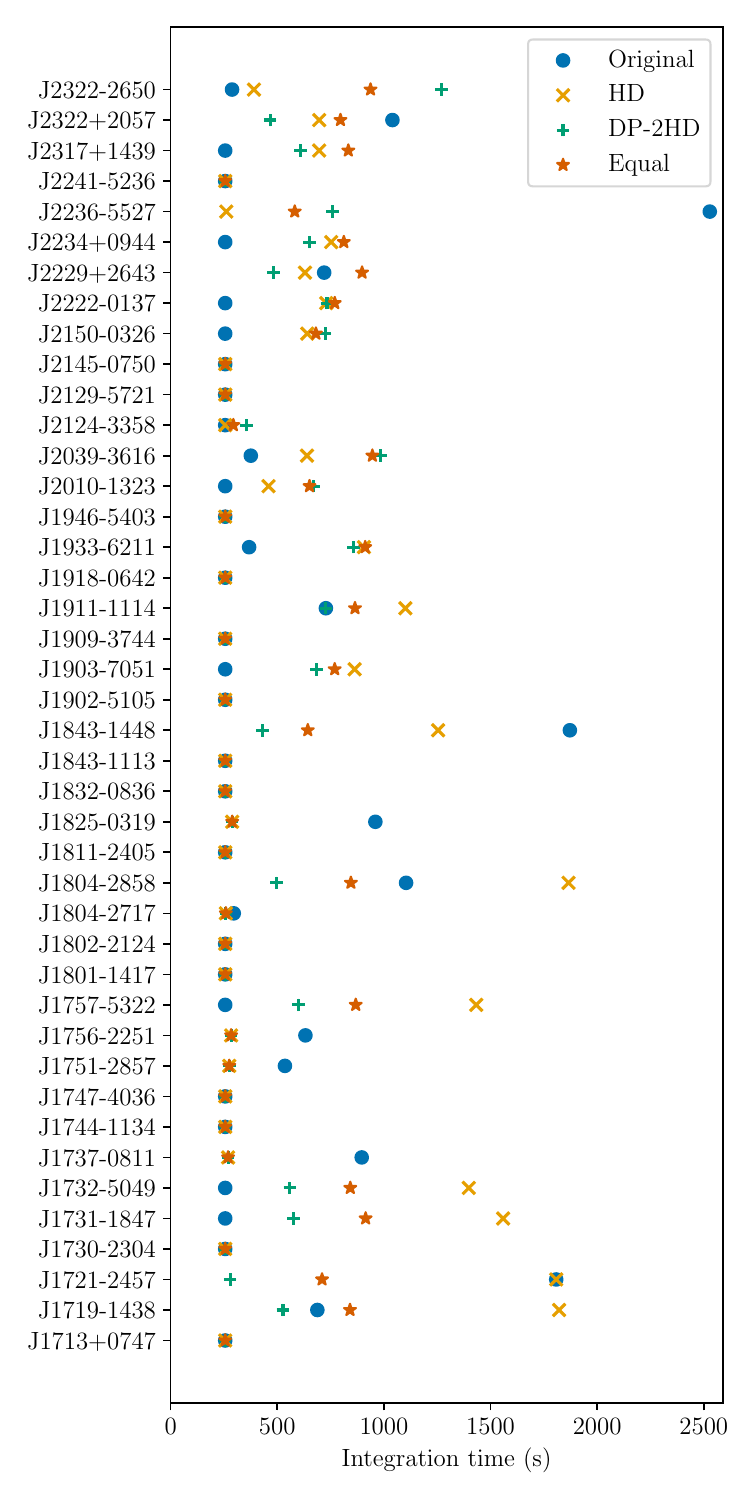}
\caption{\label{fig:obsTimes}
Integration times for each pulsar sorted in ascending right ascension from the bottom-left. 
Original times are shown as blue-circle markers. 
The other markers show integration times computed using different correlation functions for the time swap where Hellings-Downs (HD), dipole minus Hellings-Downs (DP-2HD), and equal are shown by orange-cross, green-plus, and red-star markers, respectively.
}
\end{figure*}

Figure~\ref{fig:obsTimes} shows the corresponding integration times for the PTAs configurations shown in Fig.~\ref{fig:snrCompare}.
Using different correlation functions changes which pulsars gain and lose time. 
In general, it is advantageous to reduce integration time for pulsars that require more than $1500\,{\rm s}$ to achieve $1\,{\rm \mu s}$ precision. 
Three pulsars in particular stand out as candidates for reduced integration time: 
J1231$-$1411,
J1843$-$1448,
and
J2236$-$5527.
All three of these pulsars have no measured values for spin, DM, or jitter noise, so they take the default values as described in Section~\ref{sec:data}. 
It is therefore likely that they are disfavoured purely due to the long integration time required to reach $1\,{\rm \mu s}$ timing precision. 
Figure~\ref{fig:obsTimes} also shows that there are $22$ pulsars where all time reallocation methods indicate that it is advantagous to increase the integration time. 
However, these results should be considered carefully as $16$ of these pulsars have no measured spin, DM, or jitter noise, and were therefore assigned minimum noise values for the purpose of this analysis. 
Their assumed low noise values may be the reason that these are favoured. 
Measurements of pulsar noise values are important for determining a reliable observing schedule strategy.

\section{Effects of additional time}
\label{sec:twentyPercent}

\begin{figure}
\includegraphics[width=0.48\textwidth]{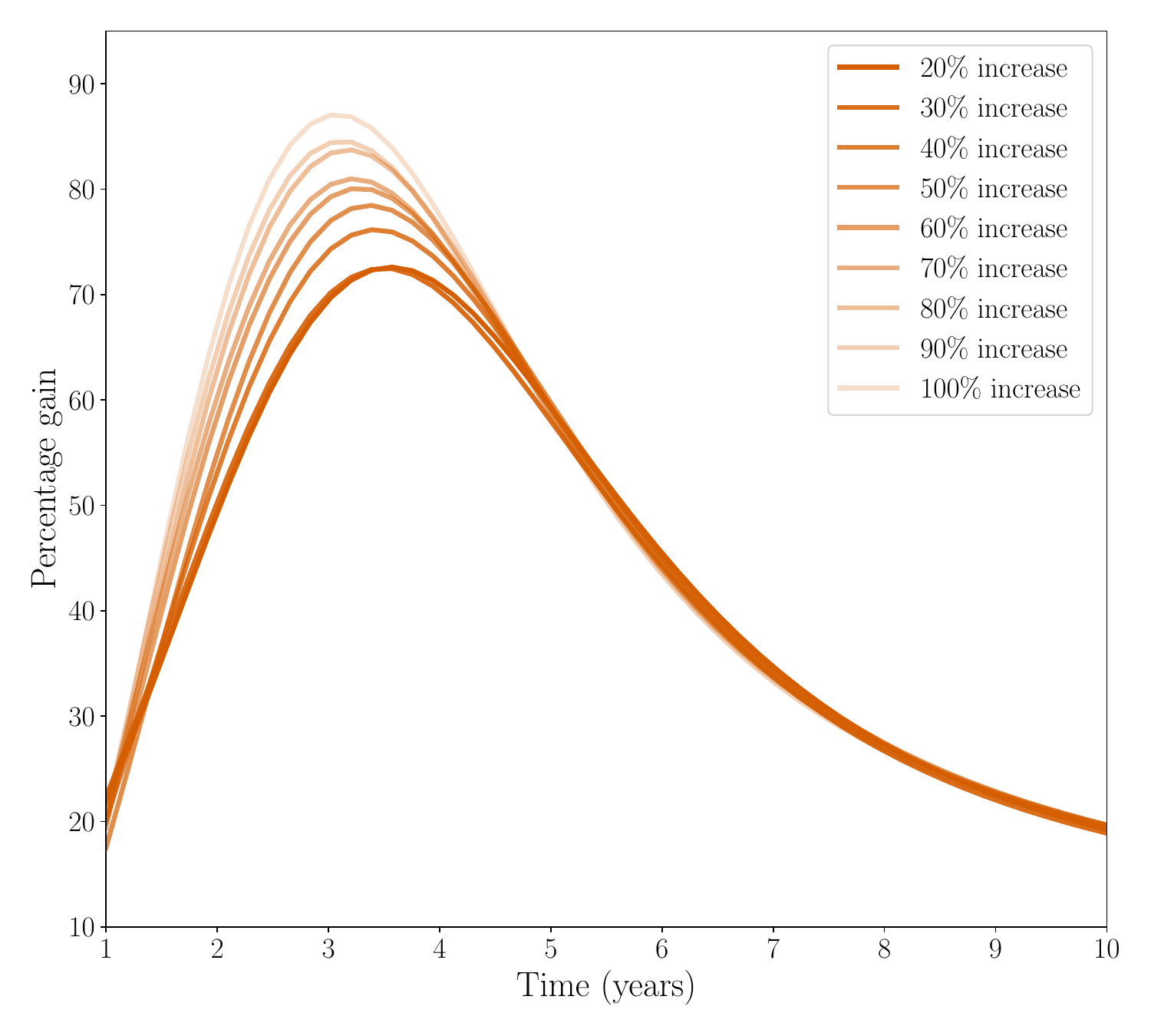}
\caption{\label{fig:extratime}
Results of the time reallocation analysis when using the starting point of assumed additional time available for the PTA. 
Each line represents a different amount of extra time allocated to MPTA starting from a $20\%$ increase (dark-red) to a $100\%$ increase (light-red).
No matter the starting point, the time-swap analysis provides a $\sim 20\%$ increase in S/N after $10\,{\rm yrs}$.
}
\end{figure}

Throughout this work, we have kept the observation time allocated to the MPTA at a constant value of $\approx \TwoWkObsTimeHrs \,{\rm hrs}$ per two weeks. 
Here we investigate how much S/N improvement could be made if the MPTA time was increased by some percentage, $q$. 
The starting point for the time-swap analysis is to increase each pulsar's integration time by this percentage so that $T_{{\rm obs},I} \rightarrow q T_{{\rm obs},I}$. 
As expected, an increase in observation time allocation gives an immediate boost to the S/N. 
For example, if there is a $20\%$ ($q=1.2$) increase in observing time (to $\approx 14\,{\rm hrs}$ every two weeks), the corresponding S/N after $10\,{\rm yrs}$ is $\approx 5\%$ higher than the current observation strategy.

As in Sections~\ref{sec:timeshuffle} and~\ref{sec:results}, we can apply the time-swap analysis to the PTA configuration using the several options for extra time ranging from an additional $20\%$ ($\Sigma_I T_{{\rm obs}}\approx 14\,{\rm hrs}$) to an additional $100\%$ ($\Sigma_I T_{{\rm obs}}\approx 23\,{\rm hrs}$). 
In this case, we only use the Equal correlation function.

The results are shown in Fig.~\ref{fig:extratime}. 
Note that the percentage gains shown here compare against the already boosted S/N of an array with extra time (as described above). 
Each of the lines inf Fig.~\ref{fig:extratime} represents a different amount of extra time devoted to PTA observations from $20\%$ in dark red to $100\%$ in light red. 
The results show that, whilst the initial gains in S/N can differ over the first $4\,{\rm yrs}$, the resulting S/N gain after $10\,{\rm yrs}$ is always around the $20\%$ mark. 
From this analysis we see that no matter the starting point, it can always be beneficial to run an analysis such as the time-swap method to find S/N gains.

\section{Conclusions}
\label{sec:conclusions}

The MPTA is capable of timing large numbers of pulsars to high precision. 
This investigation into the impact of small adjustments to the observing schedule assumes a simplified PTA, with constant cadence and minimal variation in anticipated precision, however it can provide useful insights into where gains can be made. 
Using the current allocated observing time of $\TwoWkObsTimeHrs\,{\rm hrs}$ per two-week block, we find a possible gain of $\approx 18\%$ in S/N. 
If the observation time allocated to the MPTA is increased by $20\%$, we find that the S/N increases by $\approx 27\%$ over the original observing strategy through a combination of the immediate boost in S/N from additional observing time and the time-swap analysis. 

We note that these results are heavily dependent on the noise properties assumed for each pulsar and those pulsars for which the noise properties are assumed rather than measured are less reliable. 
Continuing to measure and refine the noise properties of pulsars will improve the reliability of calculations such as the time-swap analysis demonstrated here.
Nonetheless, the results presented here can provide possible clues to S/N gains through minor changes to the observing schedule.

\section*{Acknowledgements}

The MeerKAT telescope is operated by the South African Radio Astronomy Observatory, which is a facility of the National Research Foundation, an agency of the Department of Science and Innovation.
We acknowledge the Wurundjeri People of the Kulin Nation as the Traditional Owners of the land where this work was primarily carried out. 
Computations were performed on the OzSTAR national facility at Swinburne University of Technology. The OzSTAR program receives funding in part from the Astronomy National Collaborative Research Infrastructure Strategy (NCRIS) allocation provided by the Australian Government. 
HM is grateful to Alberto Vecchio, Paul Brook, Christopher Moore, and Pratyasha Gitika for useful discussions and feedback on the manuscript.
This research is supported by the Australian Research Council Centre of Excellence for Gravitational Wave Discovery (OzGrav) (project number  CE170100004), including the OzGrav COVID funding scheme.
HM acknowledges support from the UK Space Agency, Grant No. ST/Y004922/1 and ST/V002813/1 and ST/X002071/1.
MK acknowledges significant support from the Max-Planck Society (MPG) and the MPIfR contribution to the PTUSE hardware.
AP acknowledges funding from the INAF Large Grant 2022 ``GCjewels'' (P.I. Andrea Possenti) approved with the Presidential Decree 30/2022.
This work was supported in part by the ``Italian Ministry of Foreign Affairs and International Cooperation'', grant number ZA23GR03, under the project ``\emph{RADIOMAP- Science and technology pathways to MeerKAT+: the Italian and South African synergy}''.
This work has made use of 
\texttt{astropy}~\citep{astropy:2018},
\texttt{numpy}~\citep{numpyHarris2020array},
\texttt{scipy}~\citep{scipy2020SciPy-NMeth},
and \texttt{matplotlib}~\citep{matplotlibHunter:2007}, 
\texttt{hasasia}~\citep{Hazboun2019Hasasia}.

\section*{Data Availability}
The scripts and data used for this work can be found at \url{https://github.com/hannahm8/ptasensitivity}. 

\bibliographystyle{mnras}
\bibliography{ptabib} 

\begin{thebibliography}{}
\makeatletter
\relax
\def\mn@urlcharsother{\let\do\@makeother \do\$\do\&\do\#\do\^\do\_\do\%\do\~}
\def\mn@doi{\begingroup\mn@urlcharsother \@ifnextchar [ {\mn@doi@}
  {\mn@doi@[]}}
\def\mn@doi@[#1]#2{\def\@tempa{#1}\ifx\@tempa\@empty \href
  {http://dx.doi.org/#2} {doi:#2}\else \href {http://dx.doi.org/#2} {#1}\fi
  \endgroup}
\def\mn@eprint#1#2{\mn@eprint@#1:#2::\@nil}
\def\mn@eprint@arXiv#1{\href {http://arxiv.org/abs/#1} {{\tt arXiv:#1}}}
\def\mn@eprint@dblp#1{\href {http://dblp.uni-trier.de/rec/bibtex/#1.xml}
  {dblp:#1}}
\def\mn@eprint@#1:#2:#3:#4\@nil{\def\@tempa {#1}\def\@tempb {#2}\def\@tempc
  {#3}\ifx \@tempc \@empty \let \@tempc \@tempb \let \@tempb \@tempa \fi \ifx
  \@tempb \@empty \def\@tempb {arXiv}\fi \@ifundefined
  {mn@eprint@\@tempb}{\@tempb:\@tempc}{\expandafter \expandafter \csname
  mn@eprint@\@tempb\endcsname \expandafter{\@tempc}}}

\bibitem[\protect\citeauthoryear{{Afzal} et~al.,}{{Afzal}
  et~al.}{2023}]{NANOGravNewPhysics:2023}
{Afzal} A.,  et~al., 2023, \mn@doi [\apjl] {10.3847/2041-8213/acdc91}, \href
  {https://ui.adsabs.harvard.edu/abs/2023ApJ...951L..11A} {951, L11}

\bibitem[\protect\citeauthoryear{{Agazie} et~al.,}{{Agazie}
  et~al.}{2023a}]{NANOGravGWBAgazieEtAl:2023}
{Agazie} G.,  et~al., 2023a, \mn@doi [\apjl] {10.3847/2041-8213/acdac6}, \href
  {https://ui.adsabs.harvard.edu/abs/2023ApJ...951L...8A} {951, L8}

\bibitem[\protect\citeauthoryear{{Agazie} et~al.,}{{Agazie}
  et~al.}{2023b}]{NANOGravNoise:2023}
{Agazie} G.,  et~al., 2023b, \mn@doi [\apjl] {10.3847/2041-8213/acda88}, \href
  {https://ui.adsabs.harvard.edu/abs/2023ApJ...951L..10A} {951, L10}

\bibitem[\protect\citeauthoryear{{Agazie} et~al.,}{{Agazie}
  et~al.}{2023c}]{NANOGravMBHBInterp:2023}
{Agazie} G.,  et~al., 2023c, \mn@doi [\apjl] {10.3847/2041-8213/ace18b}, \href
  {https://ui.adsabs.harvard.edu/abs/2023ApJ...952L..37A} {952, L37}

\bibitem[\protect\citeauthoryear{{Agazie} et~al.,}{{Agazie}
  et~al.}{2024}]{AgaziEtlAl:2024}
{Agazie} G.,  et~al., 2024, \mn@doi [\apj] {10.3847/1538-4357/ad36be}, \href
  {https://ui.adsabs.harvard.edu/abs/2024ApJ...966..105A} {966, 105}

\bibitem[\protect\citeauthoryear{{Anholm}, {Ballmer}, {Creighton}, {Price}  \&
  {Siemens}}{{Anholm} et~al.}{2009}]{AnholmEtAl:2009}
{Anholm} M.,  {Ballmer} S.,  {Creighton} J. D.~E.,  {Price} L.~R.,   {Siemens}
  X.,  2009, \mn@doi [\prd] {10.1103/PhysRevD.79.084030}, \href
  {https://ui.adsabs.harvard.edu/abs/2009PhRvD..79h4030A} {79, 084030}

\bibitem[\protect\citeauthoryear{{Antoniadis} et~al.,}{{Antoniadis}
  et~al.}{2022}]{IPTADR2:2022}
{Antoniadis} J.,  et~al., 2022, \mn@doi [\mnras] {10.1093/mnras/stab3418},
  \href {https://ui.adsabs.harvard.edu/abs/2022MNRAS.510.4873A} {510, 4873}

\bibitem[\protect\citeauthoryear{{Arzoumanian} et~al.,}{{Arzoumanian}
  et~al.}{2020}]{NANOGrav12.5GWB:2020}
{Arzoumanian} Z.,  et~al., 2020, \mn@doi [\apjl] {10.3847/2041-8213/abd401},
  \href {https://ui.adsabs.harvard.edu/abs/2020ApJ...905L..34A} {905, L34}

\bibitem[\protect\citeauthoryear{{Astropy Collaboration} et~al.,}{{Astropy
  Collaboration} et~al.}{2018}]{astropy:2018}
{Astropy Collaboration} et~al., 2018, \mn@doi [\aj] {10.3847/1538-3881/aabc4f},
  \href {https://ui.adsabs.harvard.edu/abs/2018AJ....156..123A} {156, 123}

\bibitem[\protect\citeauthoryear{{Bailes} et~al.,}{{Bailes}
  et~al.}{2016}]{MeerTimeBailesEtAl:2016}
{Bailes} M.,  et~al., 2016, in MeerKAT Science: On the Pathway to the SKA.
  p.~11 (\mn@eprint {arXiv} {1803.07424}), \mn@doi{10.22323/1.277.0011}

\bibitem[\protect\citeauthoryear{{Bailes} et~al.,}{{Bailes}
  et~al.}{2020}]{MeerTimeBailesEtAl:2020}
{Bailes} M.,  et~al., 2020, \mn@doi [\pasa] {10.1017/pasa.2020.19}, \href
  {https://ui.adsabs.harvard.edu/abs/2020PASA...37...28B} {37, e028}

\bibitem[\protect\citeauthoryear{{Barausse}, {Dey}, {Crisostomi}, {Panayada},
  {Marsat}  \& {Basak}}{{Barausse} et~al.}{2023}]{BarausseEtAl:2023}
{Barausse} E.,  {Dey} K.,  {Crisostomi} M.,  {Panayada} A.,  {Marsat} S.,
  {Basak} S.,  2023, \mn@doi [arXiv e-prints] {10.48550/arXiv.2307.12245},
  \href {https://ui.adsabs.harvard.edu/abs/2023arXiv230712245B} {p.
  arXiv:2307.12245}

\bibitem[\protect\citeauthoryear{{B{\'e}csy}, {Cornish}  \&
  {Kelley}}{{B{\'e}csy} et~al.}{2022}]{BecsyEtAl:2022}
{B{\'e}csy} B.,  {Cornish} N.~J.,   {Kelley} L.~Z.,  2022, \mn@doi [\apj]
  {10.3847/1538-4357/aca1b2}, \href
  {https://ui.adsabs.harvard.edu/abs/2022ApJ...941..119B} {941, 119}

\bibitem[\protect\citeauthoryear{{Burke-Spolaor} et~al.,}{{Burke-Spolaor}
  et~al.}{2019}]{BurkeSpolaorEtAl:2019}
{Burke-Spolaor} S.,  et~al., 2019, \mn@doi [\aapr] {10.1007/s00159-019-0115-7},
  \href {https://ui.adsabs.harvard.edu/abs/2019A&ARv..27....5B} {27, 5}

\bibitem[\protect\citeauthoryear{{Chen}, {Sesana}  \& {Conselice}}{{Chen}
  et~al.}{2019}]{ChenSesanaConselice:2019}
{Chen} S.,  {Sesana} A.,   {Conselice} C.~J.,  2019, \mn@doi [\mnras]
  {10.1093/mnras/stz1722}, \href
  {https://ui.adsabs.harvard.edu/abs/2019MNRAS.488..401C} {488, 401}

\bibitem[\protect\citeauthoryear{{Chen} et~al.,}{{Chen}
  et~al.}{2021}]{ChenEtAlEPTA:2021}
{Chen} S.,  et~al., 2021, \mn@doi [\mnras] {10.1093/mnras/stab2833}, \href
  {https://ui.adsabs.harvard.edu/abs/2021MNRAS.508.4970C} {508, 4970}

\bibitem[\protect\citeauthoryear{{Chen}, {Wu}  \& {Huang}}{{Chen}
  et~al.}{2022}]{ChenWuHuangIPTADR2:2021}
{Chen} Z.-C.,  {Wu} Y.-M.,   {Huang} Q.-G.,  2022, \mn@doi [Communications in
  Theoretical Physics] {10.1088/1572-9494/ac7cdf}, \href
  {https://ui.adsabs.harvard.edu/abs/2022CoTPh..74j5402C} {74, 105402}

\bibitem[\protect\citeauthoryear{{Christy}, {Anella}, {Lommen}, {Finn},
  {Camuccio}  \& {Handzo}}{{Christy} et~al.}{2014}]{ChristyEtAl:2014}
{Christy} B.,  {Anella} R.,  {Lommen} A.,  {Finn} L.~S.,  {Camuccio} R.,
  {Handzo} E.,  2014, \mn@doi [\apj] {10.1088/0004-637X/794/2/163}, \href
  {https://ui.adsabs.harvard.edu/abs/2014ApJ...794..163C} {794, 163}

\bibitem[\protect\citeauthoryear{{Colpi} et~al.,}{{Colpi}
  et~al.}{2024}]{LISA:2024}
{Colpi} M.,  et~al., 2024, \mn@doi [arXiv e-prints]
  {10.48550/arXiv.2402.07571}, \href
  {https://ui.adsabs.harvard.edu/abs/2024arXiv240207571C} {p. arXiv:2402.07571}

\bibitem[\protect\citeauthoryear{{Cordes}}{{Cordes}}{2002}]{Cordes:2002}
{Cordes} J.~M.,  2002, in {Stanimirovic} S.,  {Altschuler} D.,  {Goldsmith} P.,
    {Salter} C.,  eds,  Astronomical Society of the Pacific Conference Series
  Vol. 278, Single-Dish Radio Astronomy: Techniques and Applications. pp
  227--250

\bibitem[\protect\citeauthoryear{{Cordes} \& {Downs}}{{Cordes} \&
  {Downs}}{1985}]{CordesDowns:1985}
{Cordes} J.~M.,  {Downs} G.~S.,  1985, \mn@doi [\apjs] {10.1086/191076}, \href
  {https://ui.adsabs.harvard.edu/abs/1985ApJS...59..343C} {59, 343}

\bibitem[\protect\citeauthoryear{{Cordes} \& {Shannon}}{{Cordes} \&
  {Shannon}}{2010}]{CordesShannon:2010}
{Cordes} J.~M.,  {Shannon} R.~M.,  2010, \mn@doi [arXiv e-prints]
  {10.48550/arXiv.1010.3785}, \href
  {https://ui.adsabs.harvard.edu/abs/2010arXiv1010.3785C} {p. arXiv:1010.3785}

\bibitem[\protect\citeauthoryear{Dechter \& Dechter}{Dechter \&
  Dechter}{1989}]{DechterDechter:1998}
Dechter A.,  Dechter R.,  1989, \mn@doi [INFORMS Journal on Computing]
  {10.1287/ijoc.1.3.181}, 1, 181

\bibitem[\protect\citeauthoryear{{Detweiler}}{{Detweiler}}{1979}]{Detweiler:1979}
{Detweiler} S.,  1979, \mn@doi [\apj] {10.1086/157593}, \href
  {https://ui.adsabs.harvard.edu/abs/1979ApJ...234.1100D} {234, 1100}

\bibitem[\protect\citeauthoryear{{EPTA Collaboration and InPTA Collaboration}
  et~al.,}{{EPTA Collaboration and InPTA Collaboration}
  et~al.}{2023a}]{EPTANoise:2023}
{EPTA Collaboration and InPTA Collaboration} et~al., 2023a, \mn@doi [\aap]
  {10.1051/0004-6361/202346842}, \href
  {https://ui.adsabs.harvard.edu/abs/2023A&A...678A..49E} {678, A49}

\bibitem[\protect\citeauthoryear{{EPTA Collaboration and InPTA Collaboration}
  et~al.,}{{EPTA Collaboration and InPTA Collaboration}
  et~al.}{2023b}]{ETPAInPTAGWBAntoniadisEtAl:2023}
{EPTA Collaboration and InPTA Collaboration} et~al., 2023b, \mn@doi [\aap]
  {10.1051/0004-6361/202346844}, \href
  {https://ui.adsabs.harvard.edu/abs/2023A&A...678A..50E} {678, A50}

\bibitem[\protect\citeauthoryear{{EPTA Collaboration and InPTA Collaboration}
  et~al.,}{{EPTA Collaboration and InPTA Collaboration}
  et~al.}{2024}]{EPTAInPTAInterp:2023}
{EPTA Collaboration and InPTA Collaboration} et~al., 2024, \mn@doi [\aap]
  {10.1051/0004-6361/202347433}, \href
  {https://ui.adsabs.harvard.edu/abs/2024A&A...685A..94E} {685, A94}

\bibitem[\protect\citeauthoryear{{Foster} \& {Backer}}{{Foster} \&
  {Backer}}{1990}]{FosterBacker:1990}
{Foster} R.~S.,  {Backer} D.~C.,  1990, \mn@doi [\apj] {10.1086/169195}, \href
  {https://ui.adsabs.harvard.edu/abs/1990ApJ...361..300F} {361, 300}

\bibitem[\protect\citeauthoryear{{Geyer} et~al.,}{{Geyer}
  et~al.}{2021}]{MeerTime1000PSR:2021}
{Geyer} M.,  et~al., 2021, \mn@doi [\mnras] {10.1093/mnras/stab1501}, \href
  {https://ui.adsabs.harvard.edu/abs/2021MNRAS.505.4468G} {505, 4468}

\bibitem[\protect\citeauthoryear{{Goncharov} et~al.,}{{Goncharov}
  et~al.}{2021a}]{GoncharovEtAl:2021}
{Goncharov} B.,  et~al., 2021a, \mn@doi [\mnras] {10.1093/mnras/staa3411},
  \href {https://ui.adsabs.harvard.edu/abs/2021MNRAS.502..478G} {502, 478}

\bibitem[\protect\citeauthoryear{{Goncharov} et~al.,}{{Goncharov}
  et~al.}{2021b}]{GoncharovEtAlOnPPTA:2021}
{Goncharov} B.,  et~al., 2021b, \mn@doi [\apjl] {10.3847/2041-8213/ac17f4},
  \href {https://ui.adsabs.harvard.edu/abs/2021ApJ...917L..19G} {917, L19}

\bibitem[\protect\citeauthoryear{{Grunthal} et~al.,}{{Grunthal}
  et~al.}{2025}]{MPTAMapsGrunthalEtAl:2025}
{Grunthal} K.,  et~al., 2025, \mn@doi [\mnras] {10.1093/mnras/stae2573}, \href
  {https://ui.adsabs.harvard.edu/abs/2025MNRAS.536.1501G} {536, 1501}

\bibitem[\protect\citeauthoryear{{Handzo}, {Christy}, {Lommen}  \&
  {Perrodin}}{{Handzo} et~al.}{2015}]{HandzoEtAl:2015}
{Handzo} E.,  {Christy} B.,  {Lommen} A.~N.,   {Perrodin} D.,  2015, arXiv
  e-prints, \href {https://ui.adsabs.harvard.edu/abs/2015arXiv151009084H} {p.
  arXiv:1510.09084}

\bibitem[\protect\citeauthoryear{Harris et~al.,}{Harris
  et~al.}{2020}]{numpyHarris2020array}
Harris C.~R.,  et~al., 2020, \mn@doi [Nature] {10.1038/s41586-020-2649-2}, 585,
  357

\bibitem[\protect\citeauthoryear{Hazboun, Romano  \& Smith}{Hazboun
  et~al.}{2019}]{Hazboun2019Hasasia}
Hazboun J.,  Romano J.,   Smith T.,  2019, \mn@doi [Journal of Open Source
  Software] {10.21105/joss.01775}, 4, 1775

\bibitem[\protect\citeauthoryear{{Hellings} \& {Downs}}{{Hellings} \&
  {Downs}}{1983}]{HellingsDowns:1983}
{Hellings} R.~W.,  {Downs} G.~S.,  1983, \mn@doi [\apjl] {10.1086/183954},
  \href {https://ui.adsabs.harvard.edu/abs/1983ApJ...265L..39H} {265, L39}

\bibitem[\protect\citeauthoryear{Hunter}{Hunter}{2007}]{matplotlibHunter:2007}
Hunter J.~D.,  2007, \mn@doi [Computing in Science \& Engineering]
  {10.1109/MCSE.2007.55}, 9, 90

\bibitem[\protect\citeauthoryear{{Jaffe} \& {Backer}}{{Jaffe} \&
  {Backer}}{2003}]{JaffeBacker:2003}
{Jaffe} A.~H.,  {Backer} D.~C.,  2003, \mn@doi [\apj] {10.1086/345443}, \href
  {https://ui.adsabs.harvard.edu/abs/2003ApJ...583..616J} {583, 616}

\bibitem[\protect\citeauthoryear{{Johnston} et~al.,}{{Johnston}
  et~al.}{2020}]{MeerTime1000PSR:2020}
{Johnston} S.,  et~al., 2020, \mn@doi [\mnras] {10.1093/mnras/staa516}, \href
  {https://ui.adsabs.harvard.edu/abs/2020MNRAS.493.3608J} {493, 3608}

\bibitem[\protect\citeauthoryear{{Kelley}, {Blecha}  \& {Hernquist}}{{Kelley}
  et~al.}{2017a}]{KelleyEtAl:2017}
{Kelley} L.~Z.,  {Blecha} L.,   {Hernquist} L.,  2017a, \mn@doi [\mnras]
  {10.1093/mnras/stw2452}, \href
  {https://ui.adsabs.harvard.edu/abs/2017MNRAS.464.3131K} {464, 3131}

\bibitem[\protect\citeauthoryear{{Kelley}, {Blecha}, {Hernquist}, {Sesana}  \&
  {Taylor}}{{Kelley} et~al.}{2017b}]{KelleyEtAlTimeToDetection:2017}
{Kelley} L.~Z.,  {Blecha} L.,  {Hernquist} L.,  {Sesana} A.,   {Taylor} S.~R.,
  2017b, \mn@doi [\mnras] {10.1093/mnras/stx1638}, \href
  {https://ui.adsabs.harvard.edu/abs/2017MNRAS.471.4508K} {471, 4508}

\bibitem[\protect\citeauthoryear{{Kramer} et~al.,}{{Kramer}
  et~al.}{2021}]{MeerTimeRelBin:2021}
{Kramer} M.,  et~al., 2021, \mn@doi [\mnras] {10.1093/mnras/stab375}, \href
  {https://ui.adsabs.harvard.edu/abs/2021MNRAS.504.2094K} {504, 2094}

\bibitem[\protect\citeauthoryear{{Lam}}{{Lam}}{2018}]{Lam:2018}
{Lam} M.~T.,  2018, \mn@doi [\apj] {10.3847/1538-4357/aae533}, \href
  {https://ui.adsabs.harvard.edu/abs/2018ApJ...868...33L} {868, 33}

\bibitem[\protect\citeauthoryear{{Lam}, {McLaughlin}, {Cordes}, {Chatterjee}
  \& {Lazio}}{{Lam} et~al.}{2018}]{LamEtAl:2018}
{Lam} M.~T.,  {McLaughlin} M.~A.,  {Cordes} J.~M.,  {Chatterjee} S.,   {Lazio}
  T.~J.~W.,  2018, \mn@doi [\apj] {10.3847/1538-4357/aac48d}, \href
  {https://ui.adsabs.harvard.edu/abs/2018ApJ...861...12L} {861, 12}

\bibitem[\protect\citeauthoryear{{Lee}, {Bassa}, {Janssen}, {Karuppusamy},
  {Kramer}, {Smits}  \& {Stappers}}{{Lee} et~al.}{2012}]{LeeEtAl:2012}
{Lee} K.~J.,  {Bassa} C.~G.,  {Janssen} G.~H.,  {Karuppusamy} R.,  {Kramer} M.,
   {Smits} R.,   {Stappers} B.~W.,  2012, \mn@doi [\mnras]
  {10.1111/j.1365-2966.2012.21070.x}, \href
  {https://ui.adsabs.harvard.edu/abs/2012MNRAS.423.2642L} {423, 2642}

\bibitem[\protect\citeauthoryear{{Lorimer} \& {Kramer}}{{Lorimer} \&
  {Kramer}}{2012}]{LorimerKramer:2012}
{Lorimer} D.~R.,  {Kramer} M.,  2012, {Handbook of Pulsar Astronomy}.
{Cambridge University Press}

\bibitem[\protect\citeauthoryear{{Middleton}, {Sesana}, {Chen}, {Vecchio}, {Del
  Pozzo}  \& {Rosado}}{{Middleton} et~al.}{2021}]{MiddEtAl:2021}
{Middleton} H.,  {Sesana} A.,  {Chen} S.,  {Vecchio} A.,  {Del Pozzo} W.,
  {Rosado} P.~A.,  2021, \mn@doi [\mnras] {10.1093/mnrasl/slab008}, \href
  {https://ui.adsabs.harvard.edu/abs/2021MNRAS.502L..99M} {502, L99}

\bibitem[\protect\citeauthoryear{{Miles} et~al.,}{{Miles}
  et~al.}{2023}]{MilesEtAl:2023}
{Miles} M.~T.,  et~al., 2023, \mn@doi [\mnras] {10.1093/mnras/stac3644}, \href
  {https://ui.adsabs.harvard.edu/abs/2023MNRAS.519.3976M} {519, 3976}

\bibitem[\protect\citeauthoryear{{Miles} et~al.,}{{Miles}
  et~al.}{2025a}]{MPTANoiseMilesEtAl:2025}
{Miles} M.~T.,  et~al., 2025a, \mn@doi [\mnras] {10.1093/mnras/stae2572}, \href
  {https://ui.adsabs.harvard.edu/abs/2025MNRAS.536.1467M} {536, 1467}

\bibitem[\protect\citeauthoryear{{Miles} et~al.,}{{Miles}
  et~al.}{2025b}]{MPTAGWMilesEtAl:2025}
{Miles} M.~T.,  et~al., 2025b, \mn@doi [\mnras] {10.1093/mnras/stae2571}, \href
  {https://ui.adsabs.harvard.edu/abs/2025MNRAS.536.1489M} {536, 1489}

\bibitem[\protect\citeauthoryear{{NANOGrav Collaboration} et~al.,}{{NANOGrav
  Collaboration} et~al.}{2015}]{NANOGrav:2015}
{NANOGrav Collaboration} et~al., 2015, \mn@doi [\apj]
  {10.1088/0004-637X/813/1/65}, \href
  {https://ui.adsabs.harvard.edu/abs/2015ApJ...813...65N} {813, 65}

\bibitem[\protect\citeauthoryear{{Parthasarathy} et~al.,}{{Parthasarathy}
  et~al.}{2021}]{ParthasarathyEtAl:2021}
{Parthasarathy} A.,  et~al., 2021, \mn@doi [\mnras] {10.1093/mnras/stab037},
  \href {https://ui.adsabs.harvard.edu/abs/2021MNRAS.502..407P} {502, 407}

\bibitem[\protect\citeauthoryear{{Pol} et~al.,}{{Pol}
  et~al.}{2021}]{PolEtAl:2021}
{Pol} N.~S.,  et~al., 2021, \mn@doi [\apjl] {10.3847/2041-8213/abf2c9}, \href
  {https://ui.adsabs.harvard.edu/abs/2021ApJ...911L..34P} {911, L34}

\bibitem[\protect\citeauthoryear{{Rajagopal} \& {Romani}}{{Rajagopal} \&
  {Romani}}{1995}]{RajagopalRomani:1995}
{Rajagopal} M.,  {Romani} R.~W.,  1995, \mn@doi [\apj] {10.1086/175813}, \href
  {https://ui.adsabs.harvard.edu/abs/1995ApJ...446..543R} {446, 543}

\bibitem[\protect\citeauthoryear{{Reardon} et~al.,}{{Reardon}
  et~al.}{2023a}]{PPTAGWReardonEtAl:2023}
{Reardon} D.~J.,  et~al., 2023a, \mn@doi [\apjl] {10.3847/2041-8213/acdd02},
  \href {https://ui.adsabs.harvard.edu/abs/2023ApJ...951L...6R} {951, L6}

\bibitem[\protect\citeauthoryear{{Reardon} et~al.,}{{Reardon}
  et~al.}{2023b}]{PPTANoiseReardonEtAl:2023}
{Reardon} D.~J.,  et~al., 2023b, \mn@doi [\apjl] {10.3847/2041-8213/acdd03},
  \href {https://ui.adsabs.harvard.edu/abs/2023ApJ...951L...7R} {951, L7}

\bibitem[\protect\citeauthoryear{{Ridolfi} et~al.,}{{Ridolfi}
  et~al.}{2021}]{MeerTimeGlobularClusterPSRs:2021}
{Ridolfi} A.,  et~al., 2021, \mn@doi [\mnras] {10.1093/mnras/stab790}, \href
  {https://ui.adsabs.harvard.edu/abs/2021MNRAS.504.1407R} {504, 1407}

\bibitem[\protect\citeauthoryear{{Roebber}}{{Roebber}}{2019}]{Roebber:2019}
{Roebber} E.,  2019, \mn@doi [\apj] {10.3847/1538-4357/ab100e}, \href
  {https://ui.adsabs.harvard.edu/abs/2019ApJ...876...55R} {876, 55}

\bibitem[\protect\citeauthoryear{{Sazhin}}{{Sazhin}}{1978}]{Sazhin:1978}
{Sazhin} M.~V.,  1978, \sovast, \href
  {https://ui.adsabs.harvard.edu/abs/1978SvA....22...36S} {22, 36}

\bibitem[\protect\citeauthoryear{{Sesana}}{{Sesana}}{2013a}]{Sesana:2013}
{Sesana} A.,  2013a, \mn@doi [Classical and Quantum Gravity]
  {10.1088/0264-9381/30/24/244009}, \href
  {https://ui.adsabs.harvard.edu/abs/2013CQGra..30x4009S} {30, 244009}

\bibitem[\protect\citeauthoryear{{Sesana}}{{Sesana}}{2013b}]{SesanaExpectedGWB:2013}
{Sesana} A.,  2013b, \mn@doi [\mnras] {10.1093/mnrasl/slt034}, \href
  {https://ui.adsabs.harvard.edu/abs/2013MNRAS.433L...1S} {433, L1}

\bibitem[\protect\citeauthoryear{{Shannon} \& {Cordes}}{{Shannon} \&
  {Cordes}}{2010}]{ShannonCordes:2010}
{Shannon} R.~M.,  {Cordes} J.~M.,  2010, \mn@doi [\apj]
  {10.1088/0004-637X/725/2/1607}, \href
  {https://ui.adsabs.harvard.edu/abs/2010ApJ...725.1607S} {725, 1607}

\bibitem[\protect\citeauthoryear{{Siemens}, {Ellis}, {Jenet}  \&
  {Romano}}{{Siemens} et~al.}{2013}]{SiemensEtAl:2013}
{Siemens} X.,  {Ellis} J.,  {Jenet} F.,   {Romano} J.~D.,  2013, \mn@doi
  [Classical and Quantum Gravity] {10.1088/0264-9381/30/22/224015}, \href
  {https://ui.adsabs.harvard.edu/abs/2013CQGra..30v4015S} {30, 224015}

\bibitem[\protect\citeauthoryear{{Speri}, {Porayko}, {Falxa}, {Chen}, {Gair},
  {Sesana}  \& {Taylor}}{{Speri} et~al.}{2023}]{SperiEtAl:2023}
{Speri} L.,  {Porayko} N.~K.,  {Falxa} M.,  {Chen} S.,  {Gair} J.~R.,  {Sesana}
  A.,   {Taylor} S.~R.,  2023, \mn@doi [\mnras] {10.1093/mnras/stac3237}, \href
  {https://ui.adsabs.harvard.edu/abs/2023MNRAS.518.1802S} {518, 1802}

\bibitem[\protect\citeauthoryear{Spiewak et~al.,}{Spiewak
  et~al.}{2022}]{SpiewakEtAl:2022}
Spiewak R.,  et~al., 2022, \mn@doi [Publications of the Astronomical Society of
  Australia] {10.1017/pasa.2022.19}, 39, e027

\bibitem[\protect\citeauthoryear{{Steinle} et~al.,}{{Steinle}
  et~al.}{2023}]{SteinleEtAl:2023}
{Steinle} N.,  et~al., 2023, \mn@doi [\mnras] {10.1093/mnras/stad2408}, \href
  {https://ui.adsabs.harvard.edu/abs/2023MNRAS.tmp.2357S} {}

\bibitem[\protect\citeauthoryear{{Sykes}, {Middleton}, {Melatos}, {Di Matteo},
  {DeGraf}  \& {Bhowmick}}{{Sykes} et~al.}{2022}]{SykesEtAl:2022}
{Sykes} B.,  {Middleton} H.,  {Melatos} A.,  {Di Matteo} T.,  {DeGraf} C.,
  {Bhowmick} A.,  2022, \mn@doi [\mnras] {10.1093/mnras/stac388}, \href
  {https://ui.adsabs.harvard.edu/abs/2022MNRAS.511.5241S} {511, 5241}

\bibitem[\protect\citeauthoryear{{Taylor}, {Vallisneri}, {Ellis}, {Mingarelli},
  {Lazio}  \& {van Haasteren}}{{Taylor} et~al.}{2016}]{TaylorEtAl:2016}
{Taylor} S.~R.,  {Vallisneri} M.,  {Ellis} J.~A.,  {Mingarelli} C.~M.~F.,
  {Lazio} T.~J.~W.,   {van Haasteren} R.,  2016, \mn@doi [\apjl]
  {10.3847/2041-8205/819/1/L6}, \href
  {https://ui.adsabs.harvard.edu/abs/2016ApJ...819L...6T} {819, L6}

\bibitem[\protect\citeauthoryear{{Tiburzi} et~al.,}{{Tiburzi}
  et~al.}{2021}]{TiburziEtAl:2021}
{Tiburzi} C.,  et~al., 2021, \mn@doi [\aap] {10.1051/0004-6361/202039846},
  \href {https://ui.adsabs.harvard.edu/abs/2021A&A...647A..84T} {647, A84}

\bibitem[\protect\citeauthoryear{{Verbiest} et~al.,}{{Verbiest}
  et~al.}{2016}]{IPTA:2016}
{Verbiest} J.~P.~W.,  et~al., 2016, \mn@doi [\mnras] {10.1093/mnras/stw347},
  \href {https://ui.adsabs.harvard.edu/abs/2016MNRAS.458.1267V} {458, 1267}

\bibitem[\protect\citeauthoryear{{Vigeland} \& {Siemens}}{{Vigeland} \&
  {Siemens}}{2016}]{VigelandSiemens:2016}
{Vigeland} S.~J.,  {Siemens} X.,  2016, \mn@doi [\prd]
  {10.1103/PhysRevD.94.123003}, \href
  {https://ui.adsabs.harvard.edu/abs/2016PhRvD..94l3003V} {94, 123003}

\bibitem[\protect\citeauthoryear{Virtanen et~al.,}{Virtanen
  et~al.}{2020}]{scipy2020SciPy-NMeth}
Virtanen P.,  et~al., 2020, \mn@doi [Nature Methods]
  {10.1038/s41592-019-0686-2}, \href {https://rdcu.be/b08Wh} {17, 261}

\bibitem[\protect\citeauthoryear{{Wang}}{{Wang}}{2015}]{WangNoiseReview:2015}
{Wang} Y.,  2015, in Journal of Physics Conference Series. p. 012019
  (\mn@eprint {arXiv} {1505.00402}), \mn@doi{10.1088/1742-6596/610/1/012019}

\bibitem[\protect\citeauthoryear{{Xu} et~al.,}{{Xu}
  et~al.}{2023}]{CPTAGWBXuEtAl:2023}
{Xu} H.,  et~al., 2023, \mn@doi [Research in Astronomy and Astrophysics]
  {10.1088/1674-4527/acdfa5}, \href
  {https://ui.adsabs.harvard.edu/abs/2023RAA....23g5024X} {23, 075024}

\bibitem[\protect\citeauthoryear{{Zic} et~al.,}{{Zic}
  et~al.}{2022}]{ZicEtAl:2022}
{Zic} A.,  et~al., 2022, \mn@doi [\mnras] {10.1093/mnras/stac2100}, \href
  {https://ui.adsabs.harvard.edu/abs/2022MNRAS.516..410Z} {516, 410}

\makeatother
\end{thebibliography}




\bsp	
\label{lastpage}
\end{document}